\pdfoutput=1

\documentclass[aps,twocolumn,a4paper]{revtex4}

\usepackage[utf8]{inputenc}

\usepackage{amsmath}
\usepackage{amsfonts}
\usepackage{amssymb}
\usepackage{amsthm}
\usepackage{graphicx}
\usepackage[T1]{fontenc}
\usepackage{ae}
\usepackage{soul}  
\usepackage{bbold}  
\usepackage{hyperref}
\usepackage{bm}
\usepackage{epsfig}

\renewcommand{\vec}[1]{\mathbf{#1}}

\newcommand{\vop}[1]{\hat{\vec{#1}}}

\newcommand{\bra}[1]{\langle#1|}
\newcommand{\ket}[1]{|#1\rangle}
\newcommand{\braket}[2]{\langle#1|#2\rangle}

\newcommand{\up}{\uparrow}
\newcommand{\down}{\downarrow}

\newcommand{\vecsigma}{\boldsymbol \sigma}
\newcommand{\rd}{\mathrm{d}}
\newcommand{\Tr}{\mathrm{Tr}}

\renewcommand{\Im}{\,\mathrm{Im}}
\renewcommand{\Re}{\,\mathrm{Re}}

\begin{document}
\date{\today}

\title{ Tunable polarization in beam-splitter based on 2D topological insulators }
\author{D. G. Rothe and E. M. Hankiewicz}
\affiliation{
Institute for Theoretical Physics and Astrophysics, University of W\"urzburg,  D-97074 W\"urzburg, Germany}


\begin{abstract}
The typical bulk model describing 2D topological insulators (TI) consists of two types of spin-orbit terms, the 
so-called Dirac  term which induces out-of plane spin polarization and the
Rashba term which induces in-plane spin polarization.
We show that for some parameters of the Fermi energy, 
the beam splitter device built on 2D TIs can achieve higher in-plane spin polarization than one built on materials described 
by the Rashba model itself.
Further, due to high tunability of the electron density
and the asymmetry of the quantum well, spin polarization in different directions can be obtained.
While in the normal (topologically trivial) regime the in-plane spin polarization would dominate, in the inverted regime the 
out-of-plane polarization is more significant not only in the band gap but also for small Fermi energies above the gap.
Further, we suggest a double beam splitter scheme, to measure in-plane spin current all electrically.
Although we consider here as an example HgTe/CdTe quantum wells, this scheme could be also promising for InAs/GaSb QWs
where the in- and out-of-plane polarization could be achieved in a single device.

\end{abstract}
\maketitle

\section{Introduction}
\label{sectionIntroduction}
One of the major goals in the field of spintronics is the generation and detection of spin currents \cite{Wolf01}.
Most of the proposals rely on external magnetic fields or ferromagnetic constituents. 
However, spin injection from a ferromagnet into a semiconductor turns out to be problematic.
Attempts to remove the Schottky barrier by doping may lead to spin dephasing at the interface, and also
conductivity mismatch turns out to be a fundamental problem \cite{Schmidt2000} for spin current injection.
There have been successful experimental attempts using vacuum tunneling \cite{Alvarado92} and ferromagnetic semiconductors \cite{Ohno99}.
Considering the impediments integrating ferromagnet-semiconductor interfaces or applied magnetic fields into the technology, 
it is desirable to find spin current injection based on all-electrical principles. 

One  possibility to generate a  spin current by all electrical means is by using spin-Hall effect (SHE)
i.e.  generation of  a spin current perpendicular to the applied field in the medium with a spin-orbit coupling.
One of the sources of a transverse spin current in this context could be a  spin-dependent scattering off impurities \cite{Dyakonov2, Hirsch99}.
Later, it has been realized that the spin Hall effect is also present in semiconductors with large spin-orbit (SO) induced band splitting \cite{Murakami03, Sinova04}.
Spin accumulation induced by the spin-Hall mechanism has been experimentally confirmed first by optical means \cite{Kato,Wunderlich}, 
and later, the SHE has  been also all-electrically detected  by combining it with the inverse SHE in metals and semiconductors \cite{Tinkham06, Bruene10}.
On the theoretical side, the interplay of different mechanisms leading to the SHE - scattering off impurities causing the skew scattering and side jump effects 
(so called extrinsic mechanisms),
and  SO splitting from the  band structure (intrinsic effects) - has been a tough problem.
The long standing debate has been only recently resolved showing that in the dc limit for the linear Rashba model the SHE due to skew scattering and side-jump
effect vanishes in the presence of the band-structure SO coupling while it is non-zero for the 2DEG with the dominant cubic Dresselhaus term \cite{TseDSarma06,Hankiewicz08,HankiewiczVignale09,BiCulcer13}.

All-electric generation of spin currents has also been proposed by pumping techniques, 
either by control over the Rashba coupling constant \cite{Governale03}, 
or directly by gates controlling tunneling constants \cite{Brosco10}.
Spin pumping, however relying on Zeeman splitting, has also been experimentally proven \cite{Watson03}.

Here, we follow a different idea, which has originally been proposed by Khodas \cite{Khodas04}.
Similar to birefringence in optics, different spin components are refracted differently at an interface where the Rashba SO coupling
is nonzero on only one side - thus allowing for a construction of a spin filter.
Based on this principle, there have also been proposals using graphene as material,
which is interesting for the study of electron optics because of its linear dispersion.
Bercioux et al. \cite{Bercioux10} have analyzed spin-dependent transmission through an infinite N-SO-N (normal- Rashba SO - normal) junction based on graphene, 
and have also predicted spin pumping based on this geometry \cite{Bercioux12}.
Further, SO terms in graphene have also been exploited to obtain spin-dependent Veselago lensing, allowing to focus the spin \cite{Asmar13}. 

Differently from these works, we choose the HgTe/CdTe quantum well (QW) as a material of interest, because of its huge intrinsic Rashba SO coupling which can be electrically tuned. 
This material has also attracted a lot of  interest because of its topologically nontrivial band structure, which manifests itself in the  
quantum spin Hall effect \cite{Kane05QSHE, Bernevig06} recently observed in several experiments \cite{Koenig07, Bruene12}. 
In this paper, we compare polarization signals generated in topologically trivial and topologically non-trivial regimes. 
We use numerical simulations to analyze generation of spin polarization and spin currents, and detection of the latter.
However, instead of directly calculating spin polarizations and currents we rather choose to calculate 
transport of a conserved quantum number which is the helicity, and will later discuss the relation to the physical spin.
Also differently from Khodas and Bercioux, we consider a more realistic setup, which is embedded in a 4-terminal device attached to leads.
Further, due to the Dirac-like form  of the Hamiltonian \cite {Bernevig06}, even for zero Rashba terms,  there is an intrinsic SO coupling already 
in the model  leading to an out of plane spin polarization \cite{Rothe10}. Therefore, the physics is more complicated  in our system, and shows a competition of different SO terms.
Interestingly, for well chosen parameters, this can even increase the achieved in-plane spin polarization. While in the normal (topologically trivial) regime the in-plane spin polarization would dominate, in the inverted regime the out-of-plane polarization is more significant not only in the band gap but also for small Fermi energies above the gap.

In the following, we give a short outline of the article.
In the Section \ref{sectionModel}, we present the model Hamiltonian for HgTe QW and show how we define and characterize the helicity polarization that is induced by Rasbha SO.
In the Section \ref{sectionInfInterface}, we discuss the simple N-SO interface, where N is defined by Dirac-like model \cite{Bernevig06} and SO denotes the Dirac-like  model with  linear and cubic 
 Rashba  SO interactions.
Since this problem corresponds to a matrix-valued 3rd order differential equation for the envelope function, 
there are issues with the proper definition of currents \cite{Li07}, proper boundary conditions \cite{Winkler93}, and spurious solutions \cite{Schuurmans85}.
We find that mapping the problem to a lattice, gives an elegant solution to all these problems. This approach is similar to an idea by Winkler \cite{Winkler93}.
We give some technical details about our wave matching approach in the Appendix \ref{AppendixWaveMat}.
Then in the Section \ref{sectionPolRealistic}, we consider an N-SO-N junction embedded in a realistic 4-terminal device.
 Here we discuss only a  linear Rashba term in SO region, as is justified on the basis of a single junction analysis.
We discuss aspects about a geometry suitable for splitting of helicity components, discuss numerical methods,
and also elaborate on some technical questions concerning the measurement of spin currents or helicity currents.
 Finally, to obtain a better understanding of the interplay between an  in-plane polarization coming from Rashba physics and an out-of-plane polarization originating from the Dirac-like physics,
 we employ an effective perturbative model.
To link the numerical analysis with possibilities of experimental detection, we first discuss a combined device with polarizer and analyzer,
in the Section \ref{sectionDetectionScheme},
and obtain a relation between helicity and the physical electron spin, in the Section \ref{sectionSpinHelicity}.
In the Appendix \ref{AppendixHelicityOp}, we give a detailed derivation and discussion of the helicity operator for the 4-band model.
Details on the S-matrix and how it is related to the helicity current are given in the Appendix \ref{AppendixObsSmat}.
Measurement and correct definition of spin currents in structures, where spin is not conserved, has raised a lot of discussion in the literature.
Therefore, we derive a generalized continuity equation including a torque term \cite{ShiNiu06}, adapted to our Hamiltonian, in the Appendix \ref{AppendixContTorque}.
This serves to underline the physical meaning of the helicity current.

\section{Model and Characterization of Polarization}
\label{sectionModel}
\begin{figure}
\includegraphics[width=0.25\textwidth]{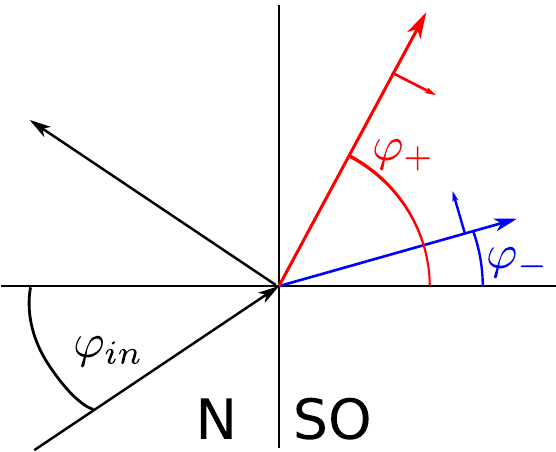} \\  (a)     
\\
\includegraphics[width=0.5\textwidth]{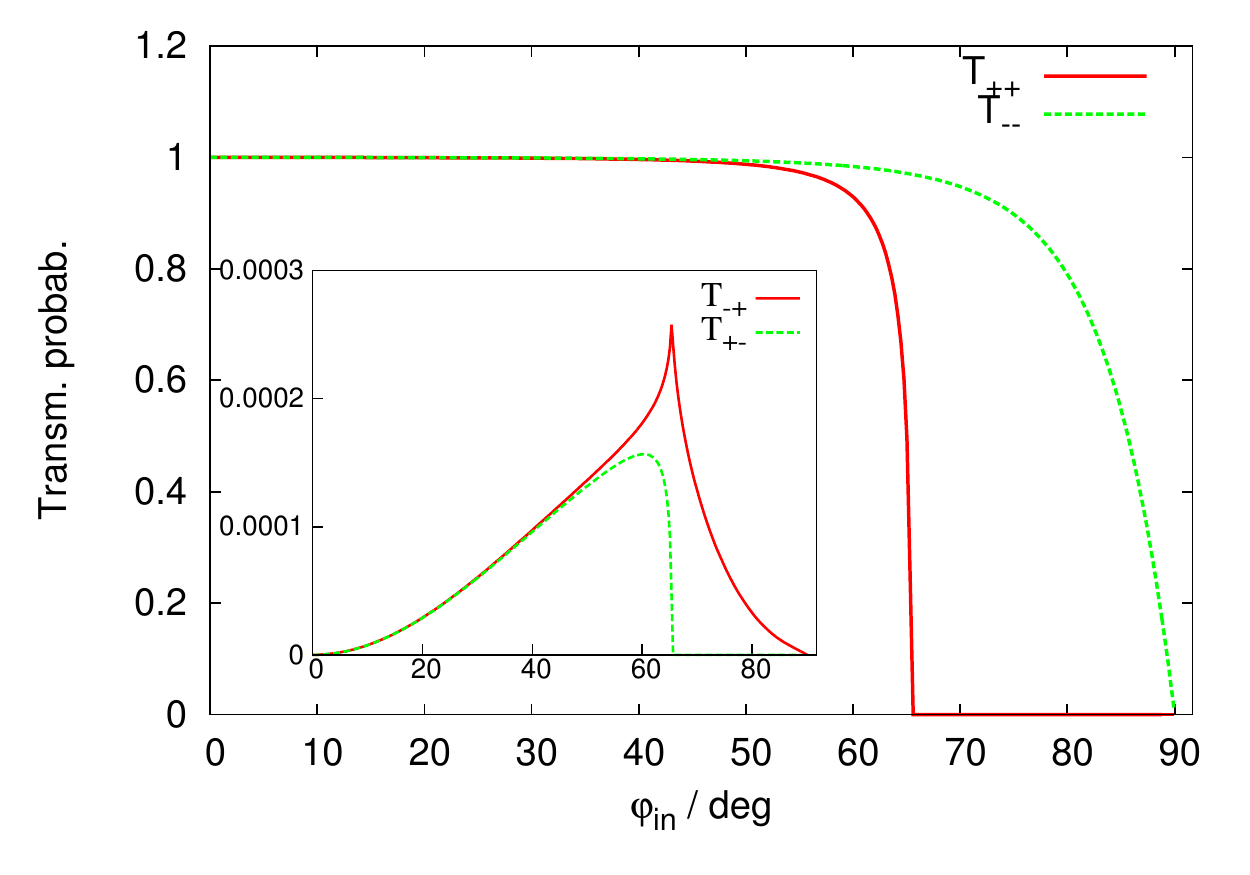} \\ (b)       
\caption{
\label{figTransmInterface}
(color online) (a) Refraction of an electron beam at an infinite N-SO interface, where the Rasbha spin-orbit Hamiltonian is given by $H_{SO}(\vec{k})$. For the same incoming angle $\varphi_{in}$, the outgoing angle $\varphi_{\pm}$ depends on the helicity $\pm 1$.
(b) The transmission probabilities  $T_{\sigma,\sigma'}$ between different helicities $\sigma,\sigma'$ as a function of the angle of incidence onto the infinite 
interface. Here the Fermi energy is $E_F = 13.8 meV$, the Rasbha SO parameters are $R = 40 meV nm$, $S = 5.4 meV nm^2$ and $T = 23 meV nm^3$ and a lattice constant 
of $a = 4.94 nm$ is used.}
\end{figure}

We consider the 4-band model for the lowest subbands of a CdTe/HgTe/CdTe quantum well (QW),
written in the basis $\ket{E+}$, $\ket{H+}$, $\ket{E-}$, $\ket{H-}$ of electron-like and heavy-hole QW subbands, 
which are angular momentum eigenstates with $S_z = \frac{1}{2},\frac{3}{2},-\frac{1}{2},-\frac{3}{2}$. The Hamiltonian \cite{Rothe10} is an extension 
of the Bernevig-Hughes-Zhang model \cite{Bernevig06, Koenig08}, and is given by
\begin{equation}
\label{hgte2deg}
H(\vec{k}) =  
\epsilon(k)I + \left(\begin{array}{cccc}
 \mathcal{M}(k)         &  A k_+              & -i R k_-        & -i S k_-^2   \\
  A k_-             &  -\mathcal{M}(k)      & i S k_-^2     & i T k_-^3  \\
  i R k_+             & -i S k_+^2                & \mathcal{M}(k)  & -A k_- \\
 i S k_+^2          & -i T k_+^3                &  -A k_+       & -\mathcal{M}(k)  
 \end{array}\right)
\end{equation}
with $\mathcal{M}(k) = M - B  k^2$, $\epsilon(k) = C - D k^2$, $k^2 = k_x^2 + k_y^2$ and $k_{\pm} = k_x \pm i k_y$.
The parameter $M$ describes the band gap and is tunable by the QW width $d$ with $M>0$ for the trivial insulator and $M<0$ for the topological insulator.
We take the parameters $A = 0.365 eV nm$, $B = -0.50 eV nm^2$, $D = -0.50 eV nm^2$ and $M = 24 meV$,
corresponding to a realistic experimental situation \cite{MuehlbauerbHankiewiczSupp} with $d = 5nm$,
in the trivial insulating regime.
For the inverted regime, we take band parameters 
$A = 0.375 eV nm$, $B = -1.120 eV nm^2$, $D = -0.730 eV nm^2$ and $M = -10 meV$,
corresponding to a QW width $d = 7nm$.
We will work with the parameters for the normal regime most of the time, comparing with the inverted regime in the Section \ref{sectionPolRealistic}.

We can decompose 
\begin{equation}
\label{HNHSO}
 H(\vec{k}) = H_N(\vec{k}) + H_{SO}(\vec{k}),
\end{equation}
where $H_N$ contains the two diagonal 2x2 blocks and $H_{SO}$ consists of the two off-diagonal 2x2 blocks,
which depend on the Rashba SO parameters $R$, $S$ and $T$. The latter are tunable by the asymmetry of the QW, e.g. by top or bottom gates.
The parameters $S$ and $T$ will give only small corrections to conduction band properties compared to $R$. 
For a given perpendicular electric field, we take the ratios $S/R$ and $T/R$ from our earlier work \cite{Rothe10}.
The Fermi energy $E_F$ is also experimentally tunable by top or bottom gates.
We will only consider $E_F$ lying in the conduction band.
In the following  $H(\vec{k})$ means the 4x4 Hamiltonian matrix, and $\hat{H}=H(\hat{\vec{k}})$ means its real space or lattice representation.

We introduce a generalization of the helicity operator to the 4-band model,
\begin{equation}
\label{helop}
h(\vec{k}) = 
\left(\begin{array}{cccc} 0 &  0  &  -ik_-/k & 0           \\
                            0 &  0  &  0         & i (k_-/k)^3 \\
                            i k_+/k &  0 &  0  &  0 \\
                            0 & -i (k_+/k)^3 & 0 & 0 
        \end{array}\right).
\end{equation}
$\hat{h} = h(\hat{\vec{k}})$ has the same symmetries as the conventional helicity operator $\vecsigma \vec{p}/p$ of the Pauli equation, i.e. 
it is parity-odd, time-reversal even, and $[\hat{H}, \hat{h}]=0$. The eigenvalues $\pm1$ of $h(\vec{k})$ are two-fold degenerate.
We will use local expectation values of $h(\vec{k})$ to define a ``helicity polarization'' which is analog to an in-plane spin polarization. 
At a given direction $\vec{k}/k$ and given $E_F$, there are two propagating eigenstates. Let $\chi$ be the normalized 4-component spinor part.
The propagating state can be chosen such that $\chi$ diagonalizes $h(\vec{k})$, i.e. $\chi^\dagger h(\vec{k}) \chi$ will be $\pm 1$.
However, $\vec{k}/k$ will not be always well-defined, 
but rather approximately known, by the direction of a lead that guides an electron beam.
Therefore, we just fix the direction $\vec{k}/k$ and define 
$h_x = h(k=k_x) = \tau_z \sigma_y$, 
where $\sigma$ Pauli matrices act on $(+,-)$-space, $\tau$ Pauli matrices  act on $(E, H)$-space, and $\sigma_0, \tau_0$ denote unit matrices.
We then use the local expectation value of $h_x$ for characterization of the helicity.
For a detailed discussion of the helicity operator, see appendix  \ref{AppendixHelicityOp}.

The local spin-z polarization, on the other hand, will be measured by the operator $\tau_0 \sigma_z$ on the 
band space (note, it is not $S_z$). If Rashba SO terms are zero ($R=S=T=0$), then $[\hat{H}, \tau_0 \sigma_z] = 0$.

\section{Infinite N-SO interface}
\label{sectionInfInterface}
In this section, we search for a maximal  value of the helicity polarization at an infinite N-SO (normal - Rashba spin orbit) interface, by means of total reflection.
We show that transmission from a mode of one helicity into a mode of the opposite helicity ( a cross-helicity term) is very small.

We consider the simple N-SO setup of Fig. \ref{figTransmInterface}a.
To the left of the interface at $x=0$, we have vanishing Rashba SO coupling terms (Hamiltonian given only by $\hat{H}_N$ of Eq. \ref{HNHSO}), and to the right 
we assume constant non-zero values for the Rashba parameters $R$, $S$, $T$ (full  $H(\vec{k})$ in Eq.\ref{HNHSO}). 
An electron beam enters from the left, with angle of incidence $\varphi_{in}$.
To the right of the interface, the components of different helicity ($\pm1$) continue at different angles $\varphi_{\pm}$. 
In Fig. \ref{figTransmInterface}a, arrows perpendicular to the momentum visualize the helicity, 
by the analogy to the direction of quantized spin in a 2DEG system with Rashba SO.

The N-SO interfaces will be the building block of the more realistic setup of the next section, compare Fig. \ref{setup}.
Compared to the analysis of Khodas  \cite{Khodas04} who also considered an N-SO interface, our model Hamiltonian is more complicated because 
it contains more bands and in particular, non-linear Rashba SO terms.
Our approach can also be used for SO-SO interfaces with different Rashba SO parameters to the left and right, but for clarity, 
we restrict ourselves to the N-SO case.

The model Hamiltonian is
\begin{align}
\hat{H} = \hat{H}_N + \theta(x) \hat{H}_{SO} \theta(x)
\end{align}
with $\theta(x) = \left\{ \begin{array}{cc} 1 & x>0 \\ 0 & x<0 \end{array} \right.$.
The exact form of the symmetrization of the  non-commuting parts $\theta(x)$ and momentum-dependent $\hat{H}_{SO}$ 
will have influence on the sharpness of the interface on the scale of a lattice constant.
However, this is a detail that has no strong influence on the resulting transmission coefficients. The chosen symmetrization gives the sharpest possible interface,
sharper than e.g. $\frac{1}{2} \left\{ \theta(x), \hat{H}_{SO} \right\}$.

The outgoing beam directions and the critical angle of total reflection of the $+$ helicity component are 
all fixed by the conserved quantities $E_F$ and $k_y$. 
It is easy to show that at the interface, cross-helicity transmission and reflection probabilities are zero for waves 
entering perpendicular to the interface ($\varphi_{in} = 0$).
(For this, we show that in the subspace of eigenstates with $k_y=0$, the space dependent Hamiltonian $\hat{H}$ and $\hat{h}$ can be diagonalized simultaneously).
For a given Fermi energy, the critical angle $\varphi_c$ can be already found from the dispersions, 
without performing any wave matching.
Let's call $E_{E,+}(k)$ the dispersion of the SO region with positive helicity, and $E_{E,0}(k)$ the dispersion of the N region.
Finding $k_y$ from $E_{E,+}(k_y) = E_F$ and 
solving for $k_{0,x}$ in $E_{E,0}(\sqrt{k_{0,x}^2 + k_y^2}) = E_F$, we have $\tan \varphi_c = \frac{k_y}{k_{0,x}}$.
We note in passing, that a top gate which gives a potential step at $x=0$, could be used to modify the critical angle, but we will not make use of this option here.
In particular we want to see if cross-helicity transmission probabilities $T_{\pm\mp}$ are also small for incoming angles $\varphi_{in} \ne 0$.
The difficulty is, that upon replacing $k_x \to -i\partial_x$, we obtain a differential equation of 
third order, and exact matching conditions at the boundary $x=0$ are difficult to find.
Instead, we use a lattice approximation of $\hat{H}$, with next nearest neighbor couplings, so we can match the dispersion of the analytical model to 
Fourier components $\sin(ak_x)$, $\sin(2 a k_x)$, $\cos(a k_x)$ and $\cos(2ak_x)$, which enter the 
dispersion of the lattice model. Moreover, this method also 
automatically excludes large-$k$ spurious solutions \cite{Schuurmans85, Winkler93} that appear for the third order differential equation.
This approximation breaks rotational symmetry, but we still find quite good helicity values (the expectation value is different from $\pm1$ by less than $10^{-6}$).
Further, this method is also quite flexible, e.g. one may smoothen the interface. 
Transmission and reflection coefficients are then calculated with the equilibrium Green's function method.
For a detailed discussion of the wave matching method, we refer to appendix \ref{AppendixWaveMat}.

Figure \ref{figTransmInterface}b shows the transmission probability as a function of $\varphi_{in}$.
We see that total reflection can occur for the $+$ component.
(We assume $R,S,T \ge 0$, so the $+$ component is the one with the higher energy for the same $k$. If we switch the sign of the perpendicular electrical field, 
$R,S,T$ change signs and total reflection will occur for the $-$ component.)
For $\varphi_{in} \ne 0$, helicity is not conserved, but the non-conservation, given by the cross-helicity transmissions 
$T_{\pm\mp}$, is quite small. 
For comparison, we have carried out the same calculation, keeping only linear Rashba terms, and we find that 
the critical angle is only slightly changed, while the cross - helicity transmission is an order of magnitude smaller, and thus also negligible.
We conclude that we do not need to include other than linear Rashba terms in the following section, where we present calculations for a realistic geometry.

\section{Polarization in finite systems}
\label{sectionPolRealistic}
\begin{figure}[h!]
\includegraphics[width=0.5 \textwidth]{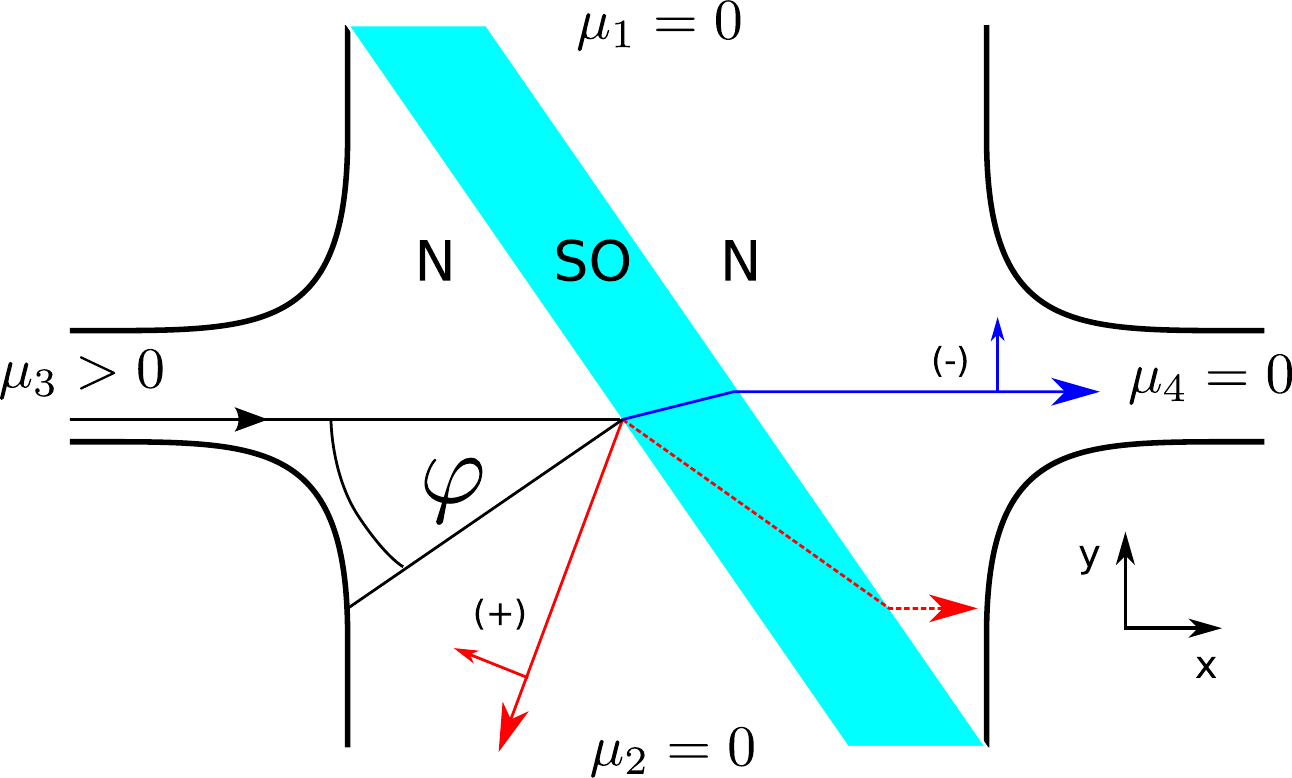}   
\caption{\label{setup} The beam splitter consists of a N-SO-N junction, tilted at angle $\varphi$, which is embedded in a 4-lead device.
If $\varphi > \varphi_c$, we expect total reflection of one helicity component
at the first N-SO interface, causing the beam leaving through the lead 4 to be polarized.  The dotted red line shows the case when $\varphi <\varphi_c$.
}
\end{figure}
\begin{figure}[h!]
\includegraphics[width=.45 \textwidth]{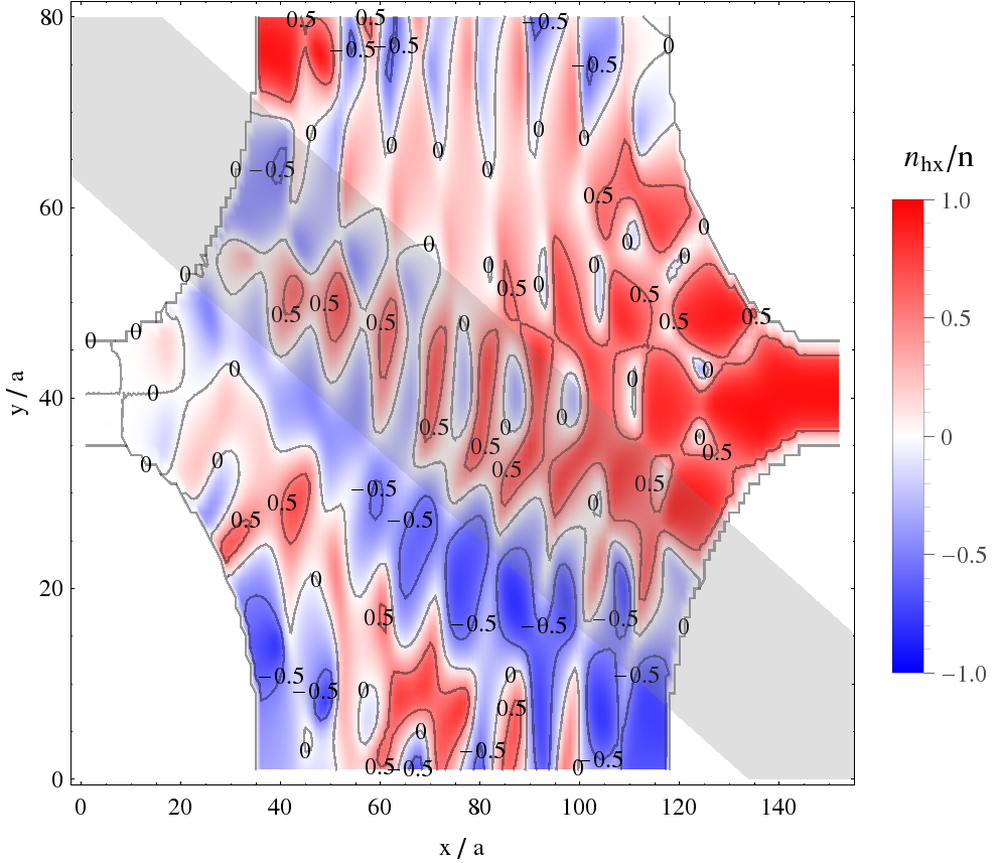}  \\(a)   
\\
\vspace{.1cm}
\includegraphics[width=.45 \textwidth]{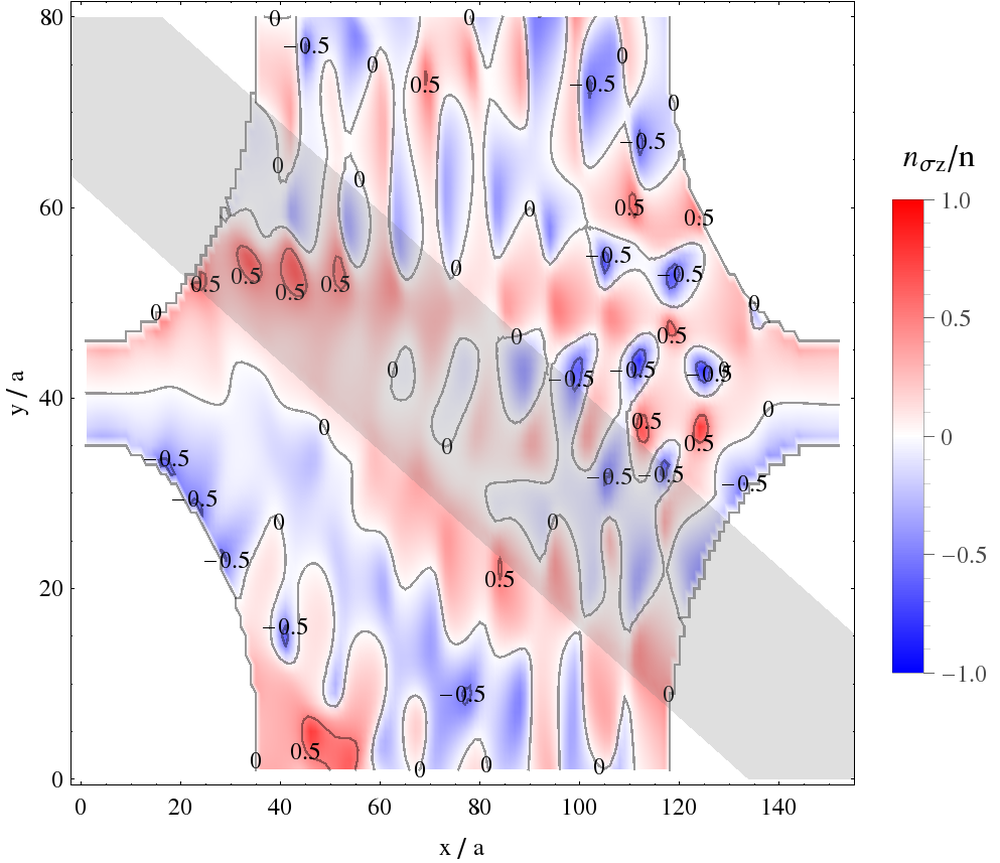} \\(b)   
\caption{\label{figdens4B} 
Local polarization plots, obtained by discretizing the 4-band model Hamiltonian on a lattice (700nm x 400nm, lattice constant $a = 4.94 nm$), 
and a geometry corresponding to Fig. \ref{setup}. Band parameters are for the normal regime.
We show the non-equilibrium response to a bias applied at the left lead.
The Fermi energy is $E_F = 0.337 t_0 = 13.8 meV$, corresponding to a peak in the polarization of the outgoing helicity current (right lead).
The linear SO parameter $R$ is nonzero only in the opaque gray area, which has a horizontal extension of $54 a$ and which is 
tilted at an angle of $\varphi = 65^\circ$, which is approximately equal to the critical angle $\varphi_c$ at the chosen $E_F$. The value of $R$ is given by $t_{SO} / t_0 = 0.1$.
(a) shows the local normalized helicity polarization assuming the direction $k=k_x$,  $n_{hx}(\vec{r})/n(\vec{r})$.
(b) shows the normalized $\sigma_z$-polarization $n_{\sigma z}(\vec{r})/n(\vec{r})$, which originates mainly from the Dirac-like physics.
}
\end{figure}
\begin{figure}[h!]
\includegraphics[width=.45 \textwidth]{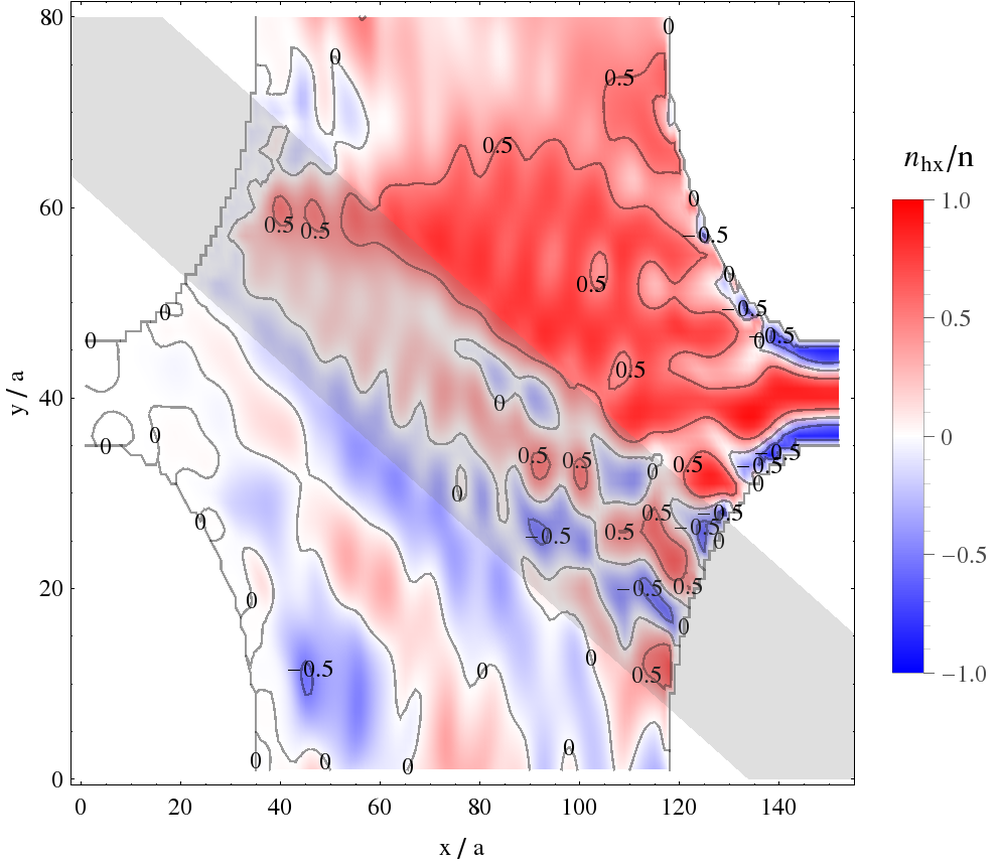}   
\\(a)
\\
\vspace{.1cm}
\includegraphics[width=.45 \textwidth]{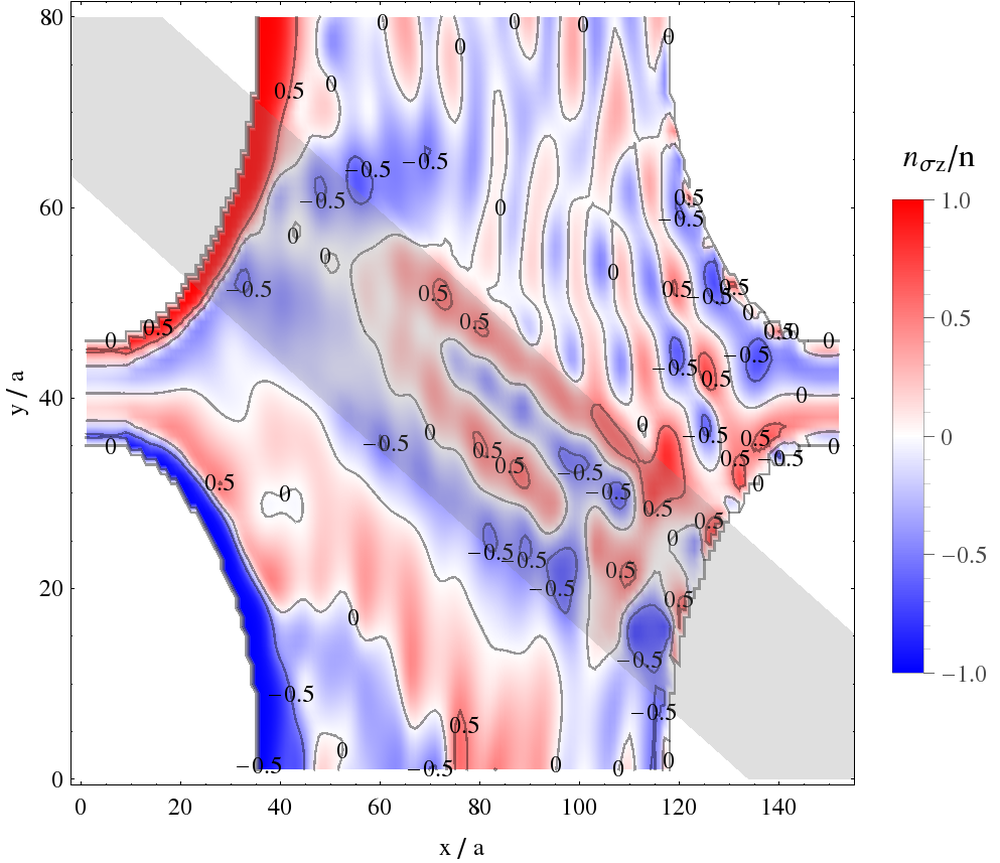}   
\\(b)
\caption{\label{figdens4Binv} 
Similarly to Figs. \ref{figdens4B}, the local normalized helicity polarization (a) and $\sigma_z$-polarization (b) are shown, but 
for the inverted ($M < 0$) regime.
 The bulk gap lies  in the range of $[-10,10] meV$.
We choose $E_F = -17.6 meV$  inside the (electron-like) valence band, corresponding to 4 propagating modes in the left/right leads.
The linear Rashba coupling in the barrier is again fixed by choosing $t_{SO}/t_0 = 0.1$.
}
\end{figure}
\begin{figure}[h]
\includegraphics[width=.5 \textwidth]{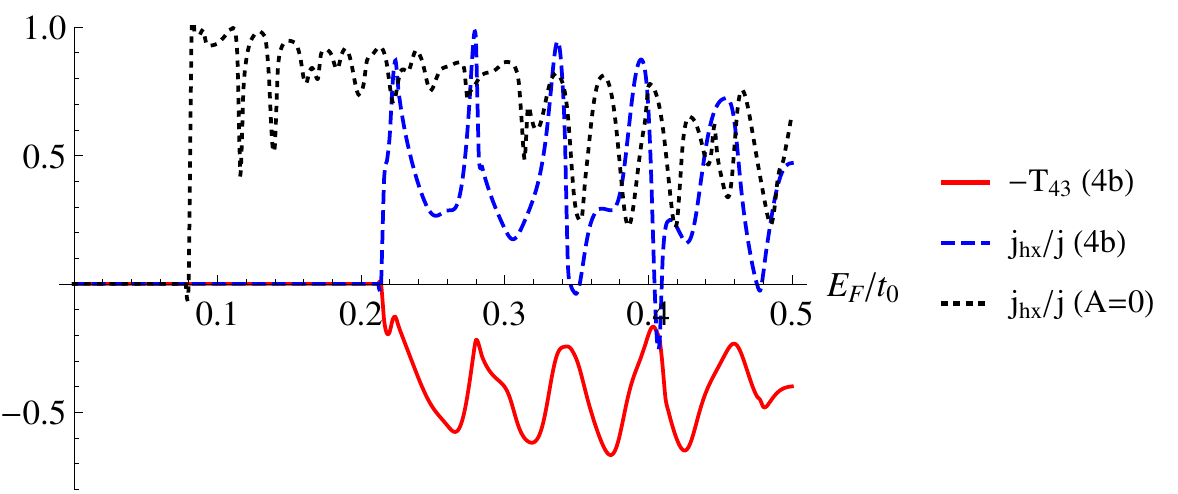}   
\\ (a) 
\\
\includegraphics[width=.5 \textwidth]{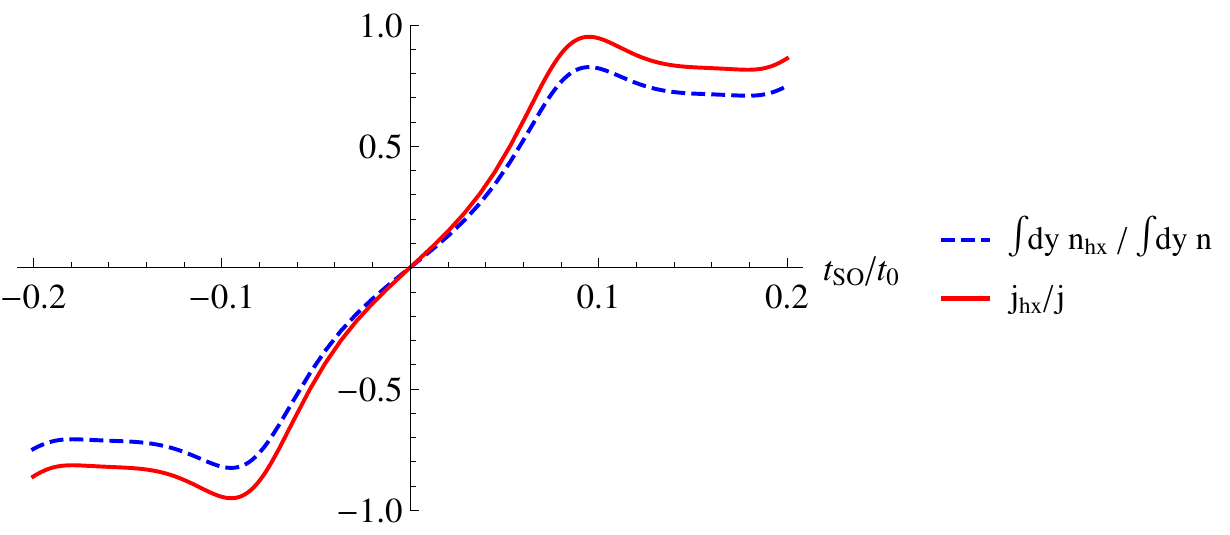}   
\\ (b)
\caption{\label{figPolOfPars}
(a) The normalized helicity current $j_{hx}/j$ at the right lead as function of the Fermi energy is shown, for the normal regime. 
Its third peak lies at $E_F = 0.337 t_0$ (``sweet spot'').
For comparison, $j_{hx}/j$ is also plotted for 
a 2-band model obtained by setting $A=0$, where it reduces to the normalized spin-y current. 
The outgoing current at the right lead, which is proportional to $T_{43}$, is shown with flipped sign for clarity. 
(b) Comparison of normalized helicity current and normalized average helicity polarization $\int dy \, n_{hx} / \int dy \, n$ in the right lead.
}
\end{figure}
\begin{figure}[h]
\includegraphics[width=.5 \textwidth]{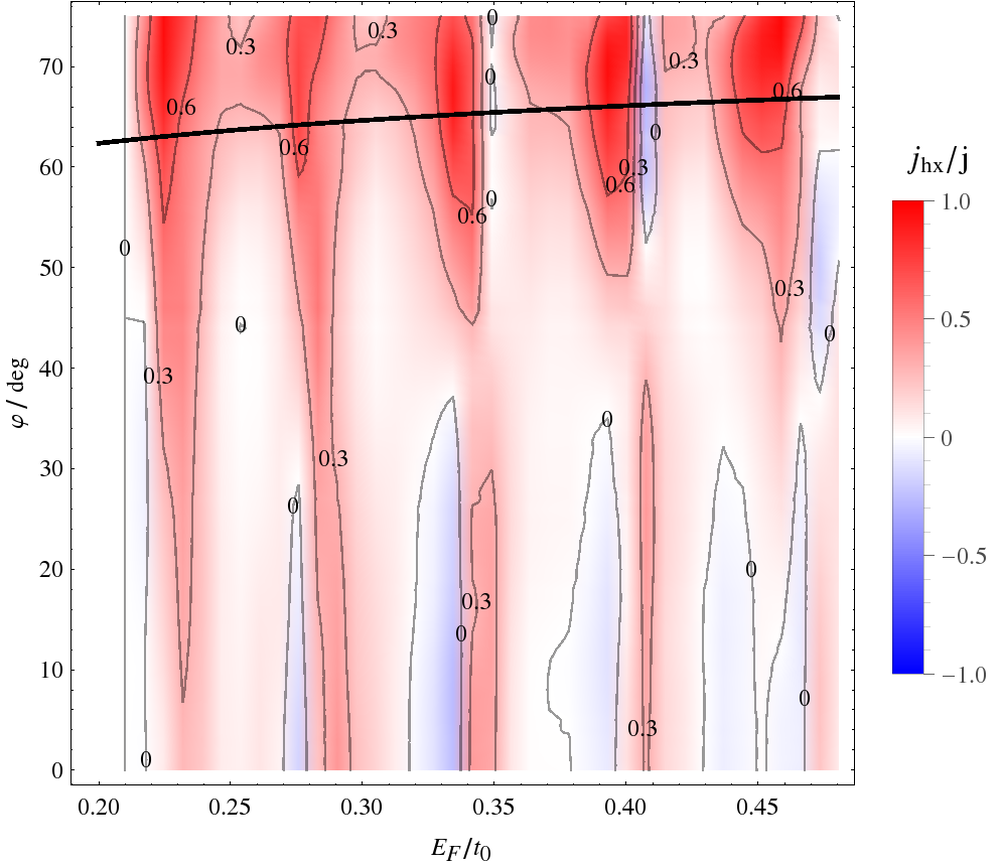}   
\caption{\label{figphiEf} Normalized  helicity current as function of Fermi energy and beam splitter tilting angle $\varphi$, for the normal regime.
Nonzero helicity current is already visible when $\varphi = 0$.
However, good polarization is only seen for $\varphi \approx \varphi_c$ or greater. 
The thick line shows $\varphi_c(E_F)$.
}
\end{figure}
In this section, we only consider the linear Rashba coupling $R$, because it is the most relevant, and put $S=T=0$.
In contrast to the N-SO interface discussed before and also analyzed by Khodas, 
we are looking for a good implementation of a spin or helicity filter in a finite geometry with attached leads. 
We give some thoughts on the geometry in the Subsection \ref{subsectionGeometry}, and  
recognize that a good spin filter device will have the form of the 4-lead setup shown in Fig. \ref{setup}.

Next, in the Subsection \ref{subsectionFormalism}, we give details about the numerical methods
used to obtain polarization and current signals, and show results.
We give a numerical comparison of the average helicity density and average helicity current in the leads and show that they are linked by a continuity equation.

Finally in the Subsection \ref{subsectionEffModel},
we provide a better understanding of the numerical results, employing an effective 2-band model.
In particular, this helps to understand the competition of the different SO terms present in the 4-band model,
which cause in-plane and out-of plane spin polarization.
We shortly comment on the validity of the model in subsection \ref{subsectionEffModelValidity}.
\subsection{Setup geometry}
\label{subsectionGeometry}
In a realistic electronic micro-device, the N-SO interfaces will have finite extension and 
the in/outgoing electron beam will be guided by the attached leads.
Figure \ref{setup} shows a tilted N-SO-N junction embedded in a 4-lead device. The electrical boundary conditions, i.e. the applied potentials $\mu_i$ at the leads,
are such that an electron beam enters from the left, at an angle of incidence $\varphi$ to the N-SO surface. 
$\varphi$ should be above the critical angle of total reflection $\varphi_c$ of the $+$ component. 
Then only the 
$-$ component traverses the SO barrier and leaves it in the same direction as it has entered. 
If $\varphi < \varphi_c$, we can still expect some helicity polarization in the right lead, because of the parallel offset of the passing $+$ beam.

The left and right leads are narrow ($49 nm$ width) and widen slowly (adiabatically).
This ensures that by the horn collimation effect (see e.g. \cite{BeenakkerHouten91} for a quantum mechanical discussion of collimation), 
the beam injected from the lead is well-directed. 
The Fermi energy is chosen low enough to have only two propagating modes in the left and right leads (not counting edge states if we are working in 
the inverted regime).
In an experimental setup, this collimation can be achieved by quantum point contacts. The upper/lower leads have to be wide
to reduce undesirable reflections. 

Our setup is invariant under rotation by $\pi$ about $\hat{z}$ ($C_2$ symmetry). Thus, a beam passing the device 
from the left will be polarized with the same efficiency as when it enters from the right. 
A simple argument shows that this symmetry is required to achieve efficient filtering of the helicity.
A device that has efficient spin/helicity filtering but not conversion, will have 
$T_{3\sigma,4\bar{\sigma}} \approx 0$, where the $\bar{\sigma}$ denotes $-\sigma$,
and $T_{p\sigma,q\sigma'}$ denotes the transmission probability from lead $q$ considering only modes with helicity $\sigma'$,
to lead $p$, with only modes with helicity $\sigma$.
Here, we use the word ``helicity'' in a loose way, assigning helicity $+$ to a transverse mode $\chi$ if the expectation value $\chi^\dagger h_x \chi$ is positive.
It will be less than $1$ because the transverse modes are not $\hat{k}_y$-eigenstates with $k_y=0$, but still, the sign is sufficient to find a simple 
description of the time reversal symmetry of the S-matrix.
If we sort in- and outgoing modes according to their helicities, time reversal symmetry makes the matrix of transmission probabilities symmetric, $T^T = T$.
Then, $\Delta T_{4,3} := T_{4+,3+} + T_{4+,3-} - T_{4-,3+} - T_{4-,3-} \approx T_{4+,3+} - T_{4-,3-} \approx \Delta T_{3,4}$.
A good efficiency in polarizing a beam entering at lead $3$ means $|\Delta T_{4,3}|$ should be large, and we see that 
this requires that a beam entering at lead $4$ and exiting at lead $3$ has to be also well polarized. 
With $C_2$ symmetry, we indeed have $T_{4+,3-} \overset{C_2}{=} T_{3+,4-} \overset{TR}{=} T_{4-,3+}$ and thus  $\Delta T_{4,3} = \Delta T_{3,4}$ holds exactly.

Khodas \cite{Khodas04} also proposes spin filtering at a single interface, based on the outgoing angle. Compared to the N-SO-N interface employed here,
spin filtering in \cite {Khodas04} works in a wider range of incoming angles, since the angle of incidence may be less than the critical angle. 
However, if leads are attached to collect the $+$ and $-$ components at different angles, this will break the $C_2$ symmetry.
Further, the outgoing beam would need re-collimation. 
In contrast, in our setup the outgoing beam has the same direction as the incoming beam, and attaching leads keeps $C_2$ symmetry.

 Note also, that with time reversal invariance and current conservation, a 2-lead spin filter is impossible with only two propagating modes ( including spin degeneracy), so the extra leads
are required. This is because, analogous to Kramer's degeneracy, the eigenvalues of the matrix $t t^\dagger$, which gives the transmission probabilities, are two-fold degenerate \cite{BardarsonPHD}.
Here, $t = S_{a,b}$ is a 2x2 submatrix of the 4x4 S-matrix, for 2 modes in each of the leads $a$,$b$.
Thus, $t t^\dagger = \gamma \mathbf 1$ is proportional to the unit matrix ($\gamma \in \mathbb R$).
Introducing the projector $P_{\sigma}$ on some unspecified spin direction, we find
$T_{b\sigma,a} = \Tr[t^\dagger P_{\sigma} t] = \gamma \Tr[P_{\sigma}] = \gamma$. The result is independent of the spin direction, thus making 
filtering of the spin impossible.  

\subsection{Formalism and polarization/current signals}
\label{subsectionFormalism}
We are interested in the helicity current at the right lead,  due to the applied bias $\mu_3$ at the left lead.
To  locally investigate some operator $\hat{O}(\vec{r})$ describing the polarization or current at position $\vec{r}=(x,y)$, we can plot 
\begin{equation}
\label{NEGFfull}
 \frac{\partial \langle\hat{O}\rangle(\vec{r}) }{\partial \mu_3}
 = \frac{1}{2\pi} \mathrm{Tr}\left[ A_3(E_F) \hat{O}(\vec{r}) \right]
\end{equation}
with the left lead contribution to the spectral density 
\begin{equation}
 A_3(E_F) = G^R \Gamma_3 G^A
\end{equation}
Here $G^{R} = (E_F - \hat{H} - \sum_{p} \Sigma_{p})^{-1}$ is the retarded Green's function, 
$G^{A} = (G^{R})^{\dagger}$ the advanced Green's function, and $\Gamma_3= i(\Sigma_3-\Sigma_3^\dagger)$ is obtained
from the left lead self-energy. The retarded self-energy of lead $p$ is given by $\Sigma_{p} = \tilde{\tau}_p (E_F + i 0^+ - \hat{H}_{p})^{-1} \tilde{\tau}_p^\dagger$,
$\tilde{\tau}_p$ is the matrix connecting the surfaces of lead $p$ and sample, and $\hat{H}_{p}$ is the Hamiltonian of the isolated semi-infinite lead.

For the operator $\hat{O}(\vec{r})$ we insert $P_{\vec{r}} h_x$ or $P_{\vec{r}} \sigma_z$  for analysis of the local helicity or spin-z polarization, respectively,
where $P_{\vec{r}} = \ket{\vec{r}}\bra{\vec{r}}$ is the projector on coordinate $\vec{r}$.
Figure \ref{figdens4B} shows 2D density plots of these signals (normalized by the local density).

For this purpose, the 4-band model Hamiltonian \eqref{hgte2deg} is discretized on a 700nm x 400nm lattice with lattice constant $a = 4.94 nm$, 
and a geometry corresponding to Figure \ref{setup}.
The linear SO parameter $R$ is nonzero only in the opaque gray area, which has a horizontal extension of $54 a$ and which is tilted at an angle of $\varphi = 65^\circ$. 
The value of $R$ is given by $t_{SO} / t_0 = 0.1$, 
where $t_{SO} = \frac{R}{2a}$ is the energy scale of Rashba SO in the lattice model and $t_0 = \frac{-B-D}{a^2}$ is the hopping energy for quadratic terms.

For the normal regime, 
we choose a Fermi energy $E_F = 0.337 t_0 = 13.8 meV$, corresponding to a peak in the polarization of the outgoing helicity current (right lead). 
The critical angle of total reflection is energy dependent, see also Figure \ref{figphiEf} where the thick 
black line shows $\varphi_c(E_F)$.
For our choice of $E_F$, we have $\varphi \approx \varphi_c$.

Figs. \ref{figdens4B}a  and \ref{figdens4Binv}a show the normalized local helicity polarization, for the topologically trivial and topologically nontrivial regimes, respectively. 
We assume  the direction $k=k_x$, i.e. we show the expectation value $n_{hx}(\vec{r}) = \frac{\delta \langle P_{\vec{r}} h_x \rangle}{\partial \mu_3} d\mu_3$,
normalized by the local (non-equilibrium) density $n(\vec{r}) = \frac{\partial \langle P_{\vec{r}} \sigma_0 \tau_0 \rangle}{\partial \mu_3} d\mu_3$.
We see that the beam is partly polarized after passing the SO barrier. 
Figs. \ref{figdens4B}b and \ref{figdens4Binv}b show the normalized $\sigma_z$-polarization, for the topologically trivial and topologically nontrivial regimes, respectively. 
We show $n_{\sigma z}(\vec{r}) = \frac{\partial \langle P_{\vec{r}} \sigma_z \tau_0 \rangle}{\partial \mu_3} d\mu_3$, again normalized by the local density.

We could also insert for $\hat{O}(\vec{r})$ the helicity current or spin-z current operators, 
$\hat{J}_{hx}(\vec{r}) = \frac{1}{4i} \{P_{\vec{r}}, \{h_x, [\hat{x}, \hat{H}]\}\}$ and 
$\hat{J}_{\sigma z}(\vec{r}) = \frac{1}{4i} \{P_{\vec{r}}, \{\tau_0 \sigma_z, [\hat{x}, \hat{H}]\}\}$.
To underline the physical meaning of such currents in transport, we have derived a general continuity equation  in the Appendix \ref{AppendixContTorque}.
It includes a torque term \cite{ShiNiu06} acting as source. However, when the average over the semi-infinite lead is taken, the torque vanishes, while the helicity current remains. 
If we are interested in the signal only in the right lead, instead of calculating the full Green's function,  it is more efficient to 
obtain the currents from the scattering matrix and expectation values of the relevant operator evaluated within the mode basis of the lead.
This way, we obtain the spin or helicity current, averaged over the semi-infinite lead. The method also works if Rashba SO terms are nonzero in the leads.
See the Appendix \ref{AppendixObsSmat} for details on the operator expectation values in terms of the S-matrix.
The scattering matrix entries $t_{pn,qm}$, i.e. the transmission amplitudes from lead $q$, mode $m$ to lead 
$p$, mode $n$, are calculated  
using a generalized Fisher-Lee relation \cite{WimmerPHD}, see eq. \eqref{tFisherLee} of  the Appendix \ref{AppendixWaveMat}.
The particle current at the right lead is proportional to the transmission probability straight through device, $T_{43} = \sum_{n,m} |t_{4n,3m}|^2$.
Actually, in this work we choose the Fermi energy low enough to have only two propagating modes (counting spin degeneracy),
and since we do not include SO terms in the leads the helicity current in the latter is conserved and identical to its average.
Note that we cannot apply the method that is usually used to calculate spin currents by introducing 
separate leads for both spin (or here, helicity) directions, because $[\hat{H}, h_x] \ne 0$.

Since the non-equilibrium helicity polarization and helicity current should be both generated by filtering out a mode of particular helicity, 
we expect signals to be qualitatively the same.
We can confirm this by the plot shown in Figure \ref{figPolOfPars}b, which shows a comparison of normalized helicity current and helicity polarization.
The normalized helicity current at the right lead is given by $j_{hx}/j$, with the definitions
$j_{hx} = \int dy \, \frac{\partial \langle \hat{J}_{hx}(x_4,y) \rangle}{\partial \mu_3}d\mu_3$ and 
$j = \int dy \, \frac{\partial \langle \hat{J}(x_4,y) \rangle}{\partial \mu_3}d\mu_3 \propto T_{43} d\mu_3$,
where $x_4$ is far in lead 4 and $\hat{J}(\vec{r})$ is defined like $\hat{J}_{hx}(\vec{r})$ with $h_x$ replaced by the unit matrix.
The normalized average helicity polarization is given by $\int dy \, n_{hx}(x_4,y) / \int dy \, n(x_4,y)$.
The signals do not oscillate as function of $x_4$, since we have only two propagating modes in the right lead, 
with the same $k_x$. We have checked that the values obtained from the full Green's function, eq. 
\eqref{NEGFfull} and values obtained from the S-matrix (Appendix \ref{AppendixObsSmat}) which needs only the surface Green's function, are almost identical. This is because evanescent modes are unimportant in the leads.
Figure \ref{figPolOfPars}a shows the normalized helicity current $j_{hx}/j$ at the right lead as 
function of the Fermi energy (blue dashed line, for the 4-band model).
The third peak lies at $E_F = 0.337 t_0$. We will use this point in the following sections, referring to it as sweet spot. Also, Figures \ref{figdens4B} are calculated for this energy. 
In Fig. \ref{figPolOfPars}a, we also plot the helicity-independent transmission probability $T_{43}$, with a flipped sign for clarity. We see that it absolut value  decrease whenever the current becomes polarized (due to the fact that the electron beam is split less electrons are transmitted to the lead 4).

In Fig. \ref{figphiEf}, we show the normalized  helicity current as function of both Fermi energy and beam splitter tilting angle $\varphi$.
Non-zero helicity current is already visible when $\varphi = 0$.
This is allowed by symmetry, since we have more than two leads, but it is not an effect of helicity-dependent refraction.
The critical angle $\varphi_c(E_F)$ is shown as thick black line.
Good polarization is only obtained for $\varphi \approx \varphi_c$ or larger, 
where we can explain the signal by the parallel displacement or total reflection of the + beam component.

\subsection{Effective 2-band model}
\label{subsectionEffModel}
In Fig. \ref{figdens4B}a the spatial map of helicity polarization $n_{hx}(\vec{r})/n(\vec{r})$ is presented (normal regime). One can see that helicity polarization oscillates as a function of the spatial coordinate.
In Fig. \ref{figdens4B}b the $\sigma_z$-polarization is shown. One can see that there is a large polarization close to the sample boundaries, 
which is well visible near the left and right leads.
Here it is important  to mention that there is non-zero out-of-plane polarization, even without the SO barrier, as a consequence of the Dirac physics.
However, as will be discussed below, the SO barrier allows to tune the degree of this polarization.

To get a better insight into these results, we will use an effective 
2-band model for just the $\ket{E\pm}$ bands in the low energy limit $A k \ll M$, that we 
already used in \cite{Rothe10, Rothe12}. Since this model is strictly valid in the normal regime, we will start with the discussion of this regime before we analyze the inverted 
(topologically non-trivial) regime.

Since electron components of the wave function are dominant, the helicity ($h_x$) polarization is approximately given by the $\sigma_y$-polarization 
which can be analyzed in the effective 2-band model. 
To allow analytical treatment, the confinement of the electrons is modeled by a confinement potential $\tau_z V(x,y)$, which replaces the 
lattice truncation for the desired geometry. This corresponds to a space-dependent band gap, so that both conduction and valence band states are confined.
The effective Hamiltonian obtained in 3rd order perturbation theory is
\begin{align}
 \label{Heffe}
\hat{H}_{e}  = \frac{k^2}{2m^*} + \underbrace{R (\vecsigma \times \hat{\vec{k}})_z}_{\text{Rashba}} 
+ \underbrace{\frac{A^2}{4 M^2} ( \nabla V \times \hat{\vec{k}})_z \sigma_z}_{ \hat{H}_D } + V(x,y) 
\end{align}
with the renormalized effective mass $m^* = (-2B-2D - \frac{A^2}{M})^{-1}$.

We expect to see a competition between the Rashba and Dirac physics in the beam splitter device. However, only the magnitude of the Rashba coupling is tunable by top and/or bottom gates.
The Dirac physics, given by $\hat{H}_D$, is due to the intrinsic SO coupling of the HgTe/CdTe material, and let us emphasize that this SO term is absent in the 2DEG model analyzed by Khodas.
In Figure \ref{figdens4B}b, the non-zero value of $n_{\sigma z}(\vec{r})$ indicates that the Dirac physics  
generates out-of plane spin polarization at the edges of the sample. Within the effective model, this 
can be explained by an anomalous velocity
\begin{align}
 \vec{v}_{an} = \frac{1}{i} [\hat{\vec{r}}, \hat{H}_D] \propto \sigma_z \left(\vec{e}_z \times \nabla V \right) 
\end{align}
which shifts spins $\uparrow,\downarrow$ into opposite directions and in the direction transverse  to the potential, which can 
be the confinement potential but also an applied potential for the electrical bias. Therefore, this effect can be interpreted as the spin-Hall effect
leading to the $\sigma_z$-polarization shown in Figure \ref{figdens4B}b, which is particularly large at the edges of the left and right leads.
In the normal regime ($M>0$), $\sigma_z$-polarization can be tuned by the change of the Fermi energy of the device or a Rashba coupling in the beam splitter part, however 
it is usually weaker than the  in-plane spin polarization. 

Further, in the signal  $n_{hx}(\vec{r})$ in  Figure \ref{figdens4B}a, we see that the helicity polarization generated by the SO barrier - 
which corresponds to an in-plane ($\sigma_y$) spin polarization of the effective 2-band model - 
is suppressed by precession of the spin around the effective $\vec{k}$-dependent magnetic field $\vec{B}_{eff} \propto \nabla V \times \vec{k}$.
This precession is also visible as sign changes of $n_{hx}$, giving a blue/red pattern along the top right sample edge. 
The phase of this precession depends on $E_F$ and therefore, good helicity polarization is obtained only for 
certain Fermi energies. This can be seen in Figures \ref{figPolOfPars}a and \ref{figphiEf}. 

Formally, we can get rid of the valence band by setting $A=0$, thus obtaining a 2-band model (2DEG) for only the conduction band
with $j_{hx}/j$  identical to the normalized spin-y current.This 2-band model should not be confused with the effective 2-band model discussed above.
In Figure \ref{figPolOfPars}a, the helicity current for the model with $A=0$ (dotted black) shows oscillations 
due to  wave interference while  different subband  quantization leads to the opening of the lowest propagating mode for lower $E_F$ than for the 4-band model.
The helicity current for the 2-band model can be now compared with the helicity current of the full 4-band model.
In the 4-band model, 
precession of the already polarized beam about  $B_{eff}$ leads to an additional structure of oscillation in comparison with 2-band model,
so that the peaks in the helicity current become more isolated in the 4-band model and the signal becomes enhanced in comparison with the 2-band model.

For completeness, we have also analyzed the helicity polarization in the 
parameter regime with band inversion ($M < 0$), where the band structure becomes topologically nontrivial.
In the inverted regime, the effective model \eqref{Heffe} does not account for the topologically protected edge states and therefore is not valid in general.
However qualitatively, we still expect to see a competition between  Rashba and  Dirac physics. In particular, the edge states are 
polarized in the spin $z$-component when Rashba SO coupling is zero, and we expect them to partially suppress the $h_x$-polarization.
In Figures \ref{figdens4Binv}, we show the normalized local helicity polarization 
$n_{hx}(\vec{r})/n(\vec{r})$ and the normalized $\sigma_z$-polarization $n_{\sigma z}(\vec{r})/n(\vec{r})$, this time for the inverted regime.
The Rashba coupling constant is chosen by the condition $t_{SO}/t_0 = 0.1$, corresponding to $R = 75 meV nm$, and is nonzero only in the tilted barrier (shown in gray).
We choose the Fermi energy $E_F = -17.6 meV$ in the (electron-like) valence band with bulk gap in the range of $[-10, 10] meV$.
With these parameters, there are actually four propagating modes in the left/right leads. However, two first modes  
are spin edge states which, although the energy is outside of the bulk gap, do not merge with the bulk, and still retain their character of being strongly localized at the sample edges.
They are almost fully polarized in $\sigma_z$ (a deviation from full polarization is due to overlap of edge states at the opposite sides of the lead).
In the $n_{\sigma z}$ plot, they can be seen to bend, following the sample edges.
Therefore, within our choice of the bias voltages, the charge signal $T_{43}$ at the right lead is zero when only edge states are propagating.
However, the next two propagating modes in the left lead pass the SO barrier from left to right, giving a nonzero $T_{43}$ at $E_F$.
Their interplay between  the bulk and the edge states also contributes to $T_{43}$, as can been seen by analyzing the S-matrix. 
The local $\sigma_z$-polarization in the right lead can be significant and is often higher than the in-plane ($h_x$) polarization.
The lead-averaged helicity current for the inverted regime is in general smaller than for the normal regime.
We do not show the polarization signals in the conduction band for the inverted regime. For small Fermi energies above the gap,
heavy hole components are dominant, and consequently 
our analysis in the Section \ref{sectionSpinHelicity} shows that in-plane spin polarization will be small since it depends on $\ket{E\pm}$ components only.
Although a relation between both polarizations is a complicated  functions of many parameters, we find that by changing $E_F$ or normal versus inverted regimes, one can tune  the ratio between  in-plane and out-of plane polarizations.

\subsection{Validity of the effective model}
\label{subsectionEffModelValidity}
We make some rough estimates to show that for our parameters, we are in the right regime 
to consider the Dirac term of the effective model as small perturbation.
We want to show that the expectation value of $\hat{H}_D = \frac{A^2}{4M^2} \left(\nabla V \times k\right)_z \sigma_z$ 
is small compared to $E_F = 13.8 meV$, which is the sweet spot. 
We may say that approximately, 
$|\langle \nabla V \times \vec{k} \rangle| < |\langle \nabla V \rangle| k_F$.
The confinement potential should be of the order of $E_F$. If $V$ is a step function with constant value zero inside, and the value $E_F$ outside of the sample region,
we can approximately say that $|\langle \nabla V \rangle| < E_F / W$, where $W$ is a length scale corresponding to the sample width,
which enters due to the normalization of the wave function.
For $W$ we enter the lead width $49 nm$ as lower bound, and we find that 
$\frac{A k_F}{2 M} \approx 0.5$ and $|\langle \hat{H}_D \rangle| < 0.1 E_F$ 
for the sweet spot, confirming the validity of our effective model.

\section{Detection scheme}
\label{sectionDetectionScheme}
\begin{figure}[h]
\includegraphics[width = .5 \textwidth]{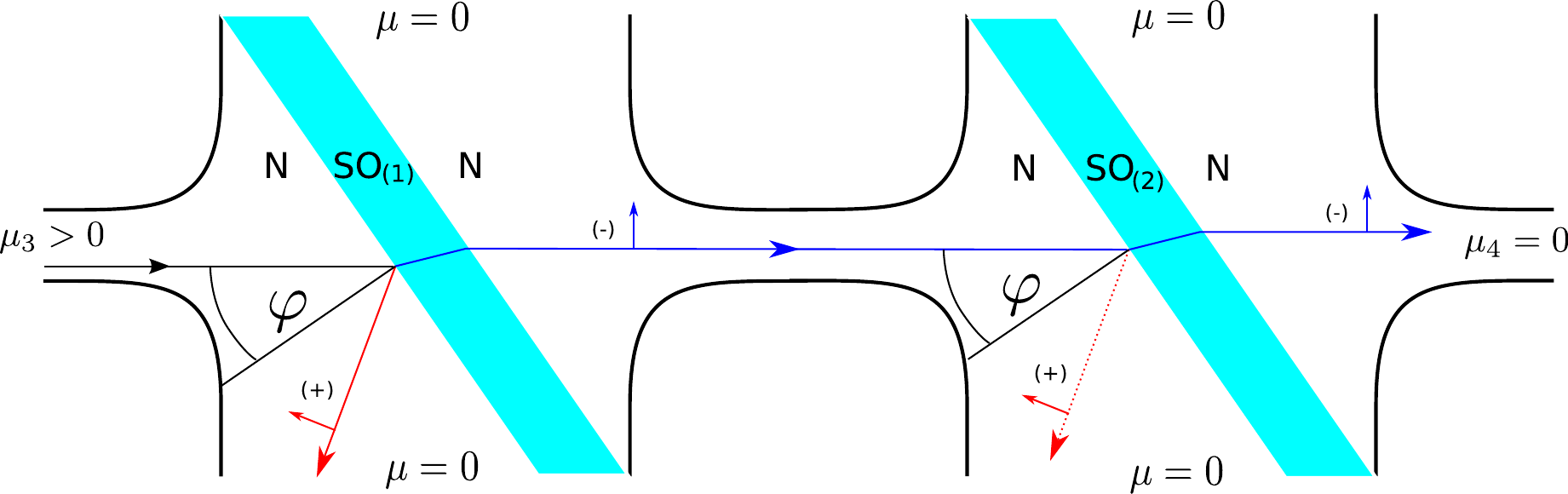}  
\caption{ \label{figDoubleDevice} A double beam splitter setup with polarizer (left) and analyzer (right) can be used to detect the helicity current all-electrically.} 
\end{figure}
\begin{figure}[h]
\includegraphics[width=.5 \textwidth]{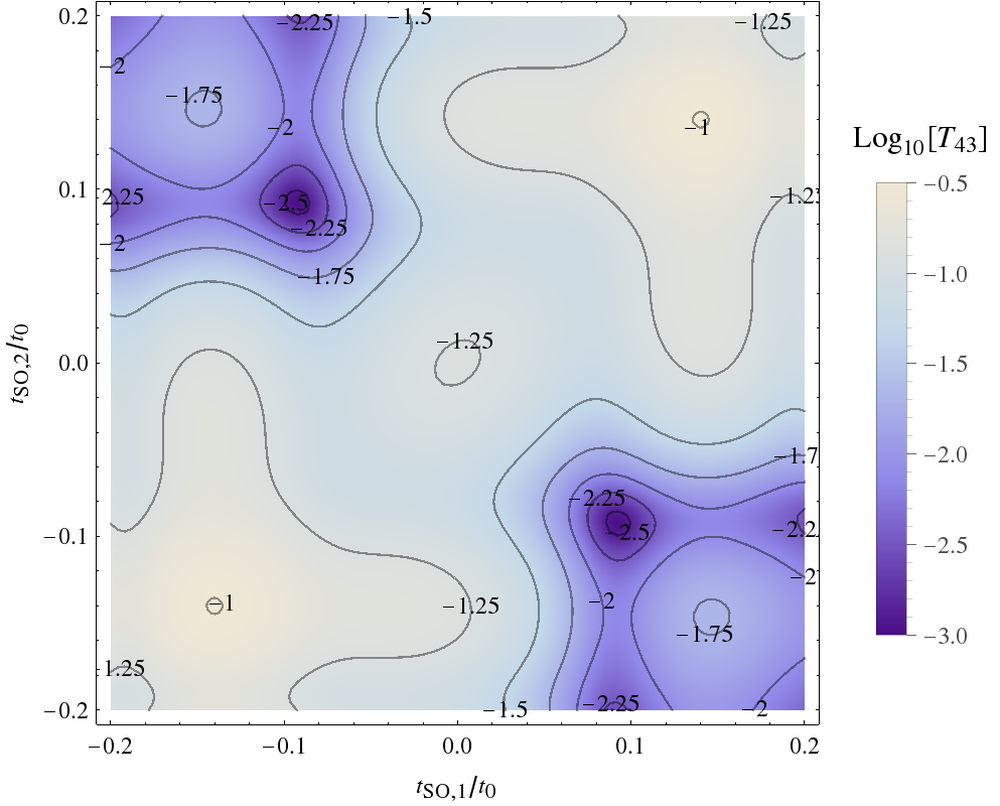}   
\\ (a)
\\ \vspace{.1cm}
\includegraphics[width=.5 \textwidth]{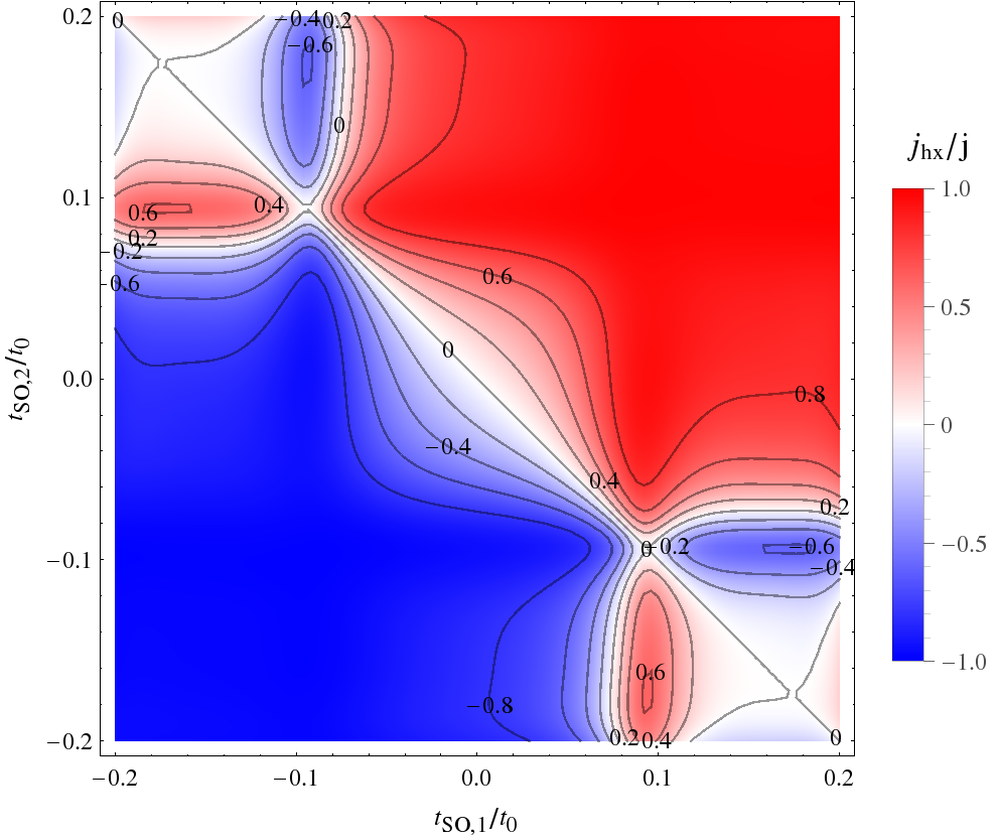}  
\\ (b)
\caption{\label{figDoubleDevice2d}
The horizontal axis parametrizes the polarizer (first device, $t_{SO,1}$) while the vertical axis parametrizes the analyzer (second device, $t_{SO,2}$).
(a) shows the transmission through the combined devices (from left to right), which is measurable all-electrically by transport. 
Here we tuned $E_F$ to the  maximal helicity current which corresponds to the maxima at $t_{SO,1}=t_{SO,2}=0.1 t_0$ and the minima at $t_{SO,1}=t_{SO,2}=-0.1 t_0$.
(b) shows the normalized helicity current in the right lead.
}
\end{figure}
Experimentally, 
spin current (or even helicity current) detection is not established so far. 
Therefore, in this section, we show that by combining two beam splitter devices with independently tunable Rashba SO parameters in the barriers, 
an all-electrical detection of the helicity current is possible.
As shown in Figure \ref{figDoubleDevice}, the first device acts as polarizer. 
The polarized beam then enters the second device. The tilting angle in both devices is assumed to be the same ($\varphi = 65^\circ$).
If the Rashba SO parameters are tuned to the same sign, the polarized beam will pass. With opposite signs, it can be blocked at the second 
device by total reflection. Thus, the transmission through the combined device, which is electrically measurable, can be used to 
prove that the current in the connecting part is polarized.

Instead of discretizing both devices on a common lattice, we can model a double beam-splitter device by combining 
the S-matrices of single beam splitters,
$S^{(i)} = \left(\begin{array}{cc} r^{(i)} & t'^{(i)} \\
                         t^{(i)} & r'^{(i)} \end{array}\right)$
($i = 1,2$).
In our case (two propagating modes in the connecting leads), $r'^{(1)}$ and $r^{(2)}$ are 2x2 matrices. 
The transmission amplitude through the combined device is \cite{Datta}
\begin{equation}
\label{tcombined}
t =  t^{(2)} ( 1 - r'^{(1)} r'^{(2)})^{-1} t^{(1)}.
\end{equation}
Note that we have to be careful with the phase definitions of the S-matrices of device 1 and 2. Normally, the 
complex phases of the S-matrix entries are undefined and thus unrelated, since the mode basis consists of asymptotic states (i.e. they are evaluated far from the scattering region). 
However here, we need to use the same phase convention 
for an outgoing mode in device 1 as for the corresponding ingoing mode in device 2 and vice versa. Otherwise the result of \eqref{tcombined} will be undefined.
If the contribution of bound states in the intermediate lead becomes negligible, this method of combining S-matrices will be exact. This means the connecting lead should 
extend over several Fermi wave lengths.
A geometric interpretation of $t$ in terms of scattering paths can be obtained by expanding in the geometric series.
The combined S-matrix can be used to find helicity currents (see appendix \ref{AppendixObsSmat}).

Fig. \ref{figDoubleDevice2d}a shows 
the transmission through a combination of polarizer and detector as in the scheme of Fig. \ref{figDoubleDevice}.
The SO parameters of the barriers ($t_{SO,1}$ vs. $t_{SO,2}$) can be tuned by two independent top gates.
$t_{SO,i} = \frac{R_i}{2a}$ is the energy scale of Rashba SO in the discretized Hamiltonian.
The horizontal axis parametrizes the polarizer (first device, $t_{SO,1}$) and the vertical axis parametrizes the analyzer (second device, $t_{SO,2}$).

The ratio of currents for parallel versus antiparallel splitters, at $t_{SO,1} = 0.1 t_0 = \pm t_{SO,2}$ is 
$0.066/0.0037 = 18$.
Of course, if we wanted to find such a high ratio for other values of $t_{SO,i}$, we would need to detune also $E_F$.
So experimentally, it would be desirable to have both a top and a bottom gate so that the Rashba parameter, controlled by the QW asymmetry, and $E_F$ could be tuned independently. 
Fig. \ref{figDoubleDevice2d}b shows the corresponding helicity current, normalized by the particle current. At $t_{SO,i} = 0.1 t_0$, we find 
the value $0.997$ (versus $0.0037$ for antiparallel splitters).
Plot \ref{figDoubleDevice2d}a is perfectly, and \ref{figDoubleDevice2d}b nearly symmetric under exchange $t_{SO,1} \leftrightarrow t_{SO,2}$.

 On the other hand, the detection of $\sigma_z$-polarization could be done by optical means like Faraday or Kerr rotation.

\section{Relation between spin and helicity}
\label{sectionSpinHelicity}
So far, we have analyzed polarization and transport in terms of the helicity operator, which we defined for this purpose. 
In this section, we want to address the obvious question, how much this observable has to do with physical (i.e. electron particle) spin polarization or currents.

For this purpose, we construct a local density matrix for the physical spin, starting from the envelope function $\psi(\vec{r})$:
\begin{align}
\label{rhosbyTrace}
 \rho^s_{i,j}(\vec{r}) = \Tr\left[\tilde{\psi}(\vec{r})\tilde{\psi}^\dagger(\vec{r}) \ket{j}_s \bra{i}_s \right]
\end{align}
Here, $\ket{i}_s = \{ \ket{\up}_s, \ket{\down}_s \}$ are basis functions of the physical electron spin.
For the envelope function, we have changed the notation from $\psi(\vec{r})$ to $\tilde{\psi}(\vec{r})$, 
emphasizing that, while $\psi(\vec{r})$ is 4-component vector depending on the in-plane coordinates $\vec{r} = (x,y)$,
$\tilde{\psi}(\vec{r})$ should be understood as tensor product of in-plane and $z$-dependent factors and orbital and spin basis functions.
The trace in \eqref{rhosbyTrace} includes an integration over $z$.
The in-plane envelope function will be expanded in components,
$\tilde{\psi}(\vec{r}) = \sum_{i=1}^4 \psi_i(\vec{r}) \ket{i}_{4b}$.
Each of the $\vec{r}$-independent basis functions $\ket{i}_{4b} = \{ \ket{E+}, \ket{H+}, \ket{E-}, \ket{H-} \}$ 
can again be expanded as $\ket{i}_{4b} = \sum_{j=1}^{6} f_{i,j}(z) \ket{j}_K$ 
in terms of new envelope function components  $f_{i,j}(z)$ that depend on the QW growth coordinate $z$, and Kane basis functions 
$\ket{j}_K = \{ \ket{\Gamma_6,\frac{1}{2}}, \ket{\Gamma_6,-\frac{1}{2}}, \ket{\Gamma_8,\frac{3}{2}}, 
\ket{\Gamma_8,\frac{1}{2}}, \ket{\Gamma_8,-\frac{1}{2}}, \ket{\Gamma_8,-\frac{3}{2}} \}$. 
In order to finally have a basis suitable for evaluation of \eqref{rhosbyTrace}, the Kane basis functions are 
in turn expanded in the orbital and spin part $\ket{j}_K = \sum_{l=S,X,Y,Z} \sum_{s=\up,\down} c^j_{l,s} \ket{l}_o \ket{s}_s$.

We introduce the convention that Pauli matrices $s_i$ act on the physical spin space. Note that $\sigma$ Pauli matrices act on $+/-$ (Kramer's) space 
and $\tau$ Pauli matrices on the E/H (QW subband) space.
For convenience, we also define the matrices $s_{\pm} = \frac{s_x \pm i s_y}{2}$ and $s_{\up/\down} = \frac{1 \pm s_z}{2}$.

As our result, we find that we can construct the space-dependent 2x2 density matrix $\rho^s(\vec{r})$ for the physical spin, 
 starting from the 4-band  wave function $\psi(\vec{r})$ and an $\vec{r}$-independent matrix,
\begin{align}
\label{rhos}
\rho^s(\vec{r}) = \psi^\dagger(\vec{r})
   \left(\begin{array}{cccc} \beta s_z + s_{\down}  &  0 & \beta s_- &0  \\
                          0      &    s_{\up} & 0 &  0 \\
                          \beta s_+ & 0 & - \beta s_z + s_{\up} & 0 \\
                          0         & 0     & 0      & s_{\down}
          \end{array}\right)
          \psi(\vec{r}).
\end{align}
Using the band parameters of our earlier paper \cite{Rothe10}, we find $\beta = 0.853$.
Next, we make use of \eqref{rhos} to find the observables in the basis of the 4-band model, that represent physical spin components,
\begin{align}
 \Tr[s_{x/y} \rho^s(\vec{r})] = \beta  \psi^\dagger(\vec{r}) \sigma_{x/y} \tau_{\up} \psi(\vec{r})
\end{align}
with $\tau_{\up} = \frac{\tau_0 + \tau_z}{2}$.
So if we compare local expectation values of $s_{y/x}$ and $h_{x/y}$, the difference is, that for the former, 
the heavy-hole components of the 4-band wave function do not contribute. This also means that generating in-plane spin polarization is not 
possible when transport is dominated by heavy holes.

We have calculated 2D density plots for $\sigma_{y} \tau_{\up}$ in the same way as for $h_x$.
Since for $E_F$ in the conduction band (normal regime), the heavy-hole wave components are small, these plots (not shown) look mostly like the $h_x$-plots, with about 10\% 
less efficiency in creating polarization (not taking into account the factor $\beta$).
Also, spin currents can be defined in terms of $\sigma_{y} \tau_{\up}$, 
but we prefer the observable $h_x$ as measure of polarization since it is related to a conserved quantity. 

For the spin-z component,
\begin{align}
 \Tr[s_z \rho^s(\vec{r})] =  \psi^\dagger(\vec{r}) \sigma_z (2 \beta \tau_{\up} - \tau_z) \psi(\vec{r})
\end{align}
For $\beta = 1$,  above observable reduces to $\sigma_z \tau_0$. We prefer the latter one as the measure of polarization since 
$[\hat{H}_N, \sigma_z \tau_0 ] = 0$.

\section{Summary}
In this paper, we analyze a beam splitter device based on 2D topological insulators. 
We find that the Dirac-like model describing these materials can lead to higher in-plane or helicity polarization than the standard model utilizing only Rashba spin-orbit interaction. 
Further, in these systems in-plane and out-of plane spin polarization can be achieved. While the trivial insulator regime ensures strong in-plane polarization, 
in the topologically non-trivial regime the interplay between edge states and bulk states induces strong out of plane polarization near the band gap.
Several important relations between spin polarization and conserved quantities like helicity polarization are established as well as 
a simple all-electrical measurement scheme for in-plane spin current using two beam splitters is proposed.
Although we focused in this paper on the parameters typical for HgTe quantum wells, the analysis presented here is also applicable to other systems 
described by the Hamiltonian of topological insulators, among them 
InAs/GaSb QWs \cite{LiuZhang08}
and 
$\text{Bi}_2\text{Se}_3$ thin films \cite{LiuZhang10}. We believe that tuning of the spin polarization from the out-of plane into the in-plane 
could be performed in one device based on InAs/GaSb QWs with top and bottom gates which change the positions of the electron and heavy-hole bands.

\section{Acknowledgements}
We acknowledge DFG Grant HA5893/1-2 within Schwerpunkt Spintronik (SPP 1285). We thank Hartmut Buhmann, Laurens Molenkamp, Mathias M\"uhlbauer, and Roland Winkler for useful discussions.
We thank Leibniz Rechenzentrum Munich for providing computing resources.
 
\clearpage
\appendix

\section{Helicity operator}
\label{AppendixHelicityOp}

\subsection{Requirements for $h(\vec{k})$}
\label{AppendixHelicityOp:req}
In order to define a helicity operator $\hat{h}$ in the 4-band model, we state some requirements that 
match the usual definition of the helicity operator, $\frac{\sigma \cdot \vec{k}}{k}$, for a spin-1/2 particle.
These requirements will make our definition unique up to an overall sign. 
$\hat{h}$ should be
\begin{itemize}
 \item 
 Hermitian,
 \item 
 time reversal symmetric: $[\mathcal{T}, \hat{h}] = 0$.
 
 For the time reversal operator $\mathcal{T} = Z K$ with $Z$ unitary and $K$ being the complex conjugation,
 we use the convention 
   $Z = -i \sigma_y \tau_0$, where $\sigma$ Pauli matrices act on $+,-$ space and $\tau$ Pauli matrices  act on $E,H$ space, and $\tau_0$ is a unit matrix. Further, $\mathcal{T} \hat{\vec{k}} \mathcal {T}^{-1} = -\hat{\vec{k}}$
 and $\mathcal{T} \vec{k} \mathcal {T}^{-1} = \vec{k}$, because $\hat{\vec{k}} = -i \nabla$ is an operator and $\vec{k}$ a real vector.

 \item 
 parity odd: $P \hat{h} P^\dagger = -\hat{h}$, 
  with $P \hat{\vec{k}} P^\dagger = -\hat{\vec{k}}$, 
    \\ $P \ket{E \pm} = -\ket{E \pm}$ and $P \ket{H \pm} = \ket{H \pm}$,
 \item
 $\hat{h}$ should have only eigenvalues $\pm 1$,
 \item
 and finally, we demand $[\hat{H}, \hat{h}] = 0$.
\end{itemize}
Note that $[\hat{H}, \hat{h}] = 0$ implies that $\hat{h}$ will also have the rotational and translational symmetry of $\hat{H}$, 
where the latter implies that we can write $\hat{h} = \int d^3 k \,  \ket{\vec{k}}h(\vec{k})\bra{\vec{k}}$.
In the following it will be shown that $h(\vec{k})$ takes the form of Eq. \eqref{helop}
of the main text.
The rotational symmetry is about the direction $\hat{z}$, which is the growth direction of the QW, and is given by 
$D_\alpha h(\phi) D_{-\alpha} = h(\phi + \alpha)$, where $\phi = \arg(k_+)$ and 
$D_\alpha = \exp(-i S_z \alpha)$ and $S_z = \mathrm{diag}(\frac{1}{2}, \frac{3}{2}, -\frac{1}{2}, -\frac{3}{2})$.

We may say that parity-oddness of $\hat{h}$ is its defining feature, 
because it corresponds to the parity-oddness of the spin-orbit terms.
We call branches of a dispersion related by time reversal, if their crossing at $k=0$ is enforced by Kramer's degeneracy.
The basic idea for the definition of $h(\vec{k})$ is to assign different signs $\pm 1$ to states at the same $\vec{k}$, if 
they are lying on dispersion branches related by time reversal symmetry. Kramer's partners will be assigned the same helicity.
If $\lambda_\pm(\vec{k})$ are the eigenvalues of $h(\vec{k})$, this is exactly what the combination of parity-oddness and time reversal 
symmetry ensures, because
$\lambda_+(\vec{k}) \overset{P}{=} - \lambda_+(- \vec{k}) \overset{\mathcal{T}}{=}  - \lambda_-(\vec{k})$.
Since SO coupling removes the degeneracy of bands related by TRS, the observable $\hat{h}$ is suitable to detect SO-related effects.

\subsection{Symmetry based derivation of $h(\vec{k})$}
\label{AppendixHelicityOp:symm}
We perform a construction of the helicity operator by symmetry, similar to the construction of $H(\vec{k})$ 
in our earlier paper \cite{Rothe10}.

The point group $T_d$ of the zinc blende structure of HgTe is reduced to the $C_{nv}$ symmetry group
by the quantum well confinement ($n$ depending on the direction of growth). In the axial approximation that we used for derivation of 
$H(\vec{k})$, we had the point group $C_{\infty v}$, and time reversal symmetry.

 The point group includes  
 reflections at a plane including $\hat{z}$, e.g. the element $\hat{C}_v$ with $\hat{C}_v (k_x,k_y) \hat{C}_v^{-1} = (-k_x,k_y)$ and 
 $\hat{C}_v\ket{E\pm} = \ket{E \mp}$, $\hat{C}_v \ket{H \pm} = \ket{H \mp}$.
 If we demand invariance of $H(\vec{k})$ under this symmetry, this enforces the parameters $A$, $R$, $S$ and $T$ to be real.
 Note that this result relies on the conventions we use for $\mathcal{T}$ and $\hat{C}_v$.

We look at the decomposition of a 4x4 matrix in terms of the Clifford algebra.
In \cite{Rothe10}, we have constructed the most general diagonal-in-$k$, rotational invariant about $\hat{z}$, time-reversal
symmetric and parity-odd Hamiltonian in the basis of $S_z$-eigenstates. The Hamiltonian
 was the part $H_{SO}$ of \eqref{hgte2deg}, which depends on the Rashba parameters $R, S, T$.
We introduce new parameters $r, s, t$ that take the places of $k R, k^2 S, k^3 T$, and obtain the most general ansatz
\begin{align}
h(\vec{k}) =  
   \left(\begin{array}{cccc}
 0               &  0            & -i r k_-/k        & -i s k_-^2/k^2   \\
 0               &  0       & i s k_-^2/k^2     & i t k_-^3/k^3  \\
  i r^* k_+/k             & -i s^* k_+^2/k^2                & 0   & 0  \\
 i s^* k_+^2/k^2          & -i t^* k_+^3/k^3                &  0    & 0
 \end{array}\right)
\end{align}
Because of the reflection symmetry $C_v$ of our system, $R$, $S$ and $T$ are real.
However, it is not yet clear that $r$, $s$ and $t$ will be real.
Evaluating $[h(\vec{k}), H(\vec{k})] = 0$, we obtain
\begin{eqnarray}
&& R \Im[r] + k S \Im[s] = 0 \\
&&  S \Im[s] + k T \Im[t] = 0  \\
&& \label{Hhcomm3}  -R s - k S t + k S r^* + k^2 T s^* = 0 \\
&& \label{Hhcomm4} 2 \mathcal{M} s + A k (r - t) = 0
\end{eqnarray}
If we say that these equations must be true independent of the parameters $R$, $S$ and 
$T$ and for all $k$, we obtain $s=0$ and $r=t$ and real. If we set $r=t=1$ in order to fix the eigenvalues to $\pm1$, we are done.
With this convention, the conduction band states with the higher energy at the same $k$, are assigned positive helicity (assuming $R,S,T>0$).

With a bit more calculation, we even do not need to assume the solution to be independent of $R$, $S$ and $T$, and we still 
obtain the same, unique result. Let us assume $R, T \ne 0$, then
\[
 \Im[r - t] = \left( \frac{S}{k T} - \frac{k S}{R} \right) \Im[s] = -\frac{2\mathcal{M}}{A k} \Im[s]
\]
Except for very special parameters of $H(\vec{k})$, we may conclude $\Im[s] = 0$, and thus $\Im[r] = 0$ and $\Im[t] = 0$.
Now that we know that $s$ is real, we may use \eqref{Hhcomm3} and \eqref{Hhcomm4},
\[
 s = \frac{k S}{k^2 T - R} (t-r) = \frac{A k}{2 \mathcal{M}} (t-r) 
\]
so we must have  $t = r$ except for very special parameters, and $s = 0$. Fixing eigenvalues to $\pm 1$, 
we obtain the result \eqref{helop}.

\subsection{Projector based derivation of $h(\vec{k})$}
\label{AppendixHelicityOp:proj}
From the derivation by symmetry, it was not yet very clear, that the different signs of helicity correspond to branches of the 
band structure that are related by time reversal symmetry. To clarify this, we give here a derivation based on projectors on 
eigenstates of $H(\vec{k})$. 

For an eigenstate $\ket{\psi_\vec{k}} = \ket{\vec{k}}\ket{\chi_+(\vec{k})}$ with spinor $\ket{\chi_+(\vec{k})}$, 
we use the time reversal operator to define a related spinor 
  $\ket{\chi_-(\vec{k})}$ at the same $\vec{k}$:
\begin{align}
\mathcal{T}\ket{\psi_{-\vec{k}}} = \ket{\vec{k}} Z \ket{\chi_+(-\vec{k})}^* =: \ket{\vec{k}} \ket{\chi_-(\vec{k})}.
\end{align}
Since we need the state at $-\vec{k}$ to find the related spinor, it is important that we have a set of eigenvectors 
that are continuous functions of $\vec{k}$. It is not possible to find eigenvectors as continuous functions 
in the complete plane of $(k_x, k_y)$, but for any given direction of $\vec{k}$, e.g. $k_y = 0$, it is possible to find eigenvectors that are continuous on this line.

We make use of the spectral representation,
\begin{equation}
 H(\vec{k}) = \sum_{\alpha=E,H} \sum_{s=+,-} \ket{\chi_{\alpha,s}(\vec{k})}\bra{\chi_{\alpha,s}(\vec{k})} E_{\alpha,s}(\vec{k}).
\end{equation}
Due to the degeneracy at $\vec{k}=0$, we have to be careful how to define eigenfunctions $\ket{\chi_\pm(\vec{k})}$ as functions of $\vec{k}$, such that they are continuous functions.
We set $E_{\alpha,\pm}(-k) = E_{\alpha,\mp}(k)$. Then, $E_{\alpha,\pm}(k)$ are differentiable functions even at the band crossing at $k=0$. We may define $\ket{\chi_{\alpha,-}(k)} = \mathcal{T}\ket{\chi_{\alpha,+}(-k)}$.
States are continuous functions on cuts $\phi = \arg(k_+) = const. + n \pi$ where $n$ takes values $0$ and $1$.

Then we can rewrite $h(\vec{k})$ using projectors on states, simply by replacing the eigenenergies with $\pm1$.
\begin{equation}
 \label{helicityInProjectors}
 h(\vec{k}) = \sum_{\alpha=E,H} s_{\alpha} \mathrm{sgn}(k_x) \left( 
 P_{\alpha,+}(\vec{k}) - P_{\alpha,-}(\vec{k}) \right)
\end{equation}
with $P_{\alpha,\pm}(\vec{k}) =  \ket{\chi_{\alpha,\pm}(\vec{k})}\bra{\chi_{\alpha,\pm}(\vec{k})}$ and signs $s_{\alpha} = \pm$ which are to be determined.
By construction, $[H(\vec{k}), h(\vec{k})] = 0$.
Under time reversal $H(\vec{k}) \to Z H^*(-\vec{k}) Z^\dagger$, we have 
$P_{\alpha,\pm}(\vec{k}) \to P_{\alpha,\mp}(\vec{k})$
so that $[\hat{h}, \mathcal{T}] = 0$.

If we assume that at $\vec{k}$ the bands are non-degenerate, this construction of $h(\vec{k})$ is unique up to the signs $s_{\alpha}$, and 
we have to show that $s_E=s_H=1$ coincides with our earlier definition.
The question of signs is equivalent to the question how we would like to label (e.g. numerically obtained) eigenstates with indices $+$ and $-$.
In particular, the relative sign $s_E/s_H$ is relevant. 
The reasonable choice is, that when considering the limit $R,S,T \to 0$, the $+$ eigenvectors should be continuously connected to eigenvectors of the 
upper left 2x2 block of $H(\vec{k})$. 
Instead of considering a continuous deformation of the parameter space of the Hamiltonian, we will use here a particle-hole symmetry to obtain an equivalent result.
As we will show in the next section, even though $H(\vec{k})$ is not particle-hole symmetric, there is an operation $P_{eh}$ (see eq \eqref{Peh}) with
\begin{align}
\label{PehProjectors}
P_{eh} \ket{\chi_{E,\pm}}\bra{\chi_{E,\pm}} P_{eh}^\dagger= \ket{\chi_{H,\pm}}\bra{\chi_{H,\pm}}.
\end{align}
$P_{eh}$ should not be confused with the parity operator $P$.
From \eqref{helicityInProjectors} it is clear that with $s_E=s_H$, we find $[h(\vec{k}), P_{eh}] =0$, while this does not hold for $s_E=-s_H$.
Since in the last section, the construction by symmetry gave a unique (parity-odd) result, for which it can be checked that $[h(\vec{k}), P_{eh}] =0$, we must have 
$s_E=s_H$ in order to fulfill parity-oddness of $h(\vec{k})$.
A more direct proof of the parity-oddness of \eqref{helicityInProjectors} would be desirable, 
but is not easy since a single projector $P_{\alpha,+}(\vec{k})$ does not have definite parity.

\subsection{Particle-hole symmetry of states}
\label{AppendixHelicityOp:phsymm}
The spectrum of $H(\vec{k})$ is not particle-hole symmetric, both because the spin-orbit terms for the E and H bands 
differently depend on $\vec{k}$, and because of $\epsilon(k)$. 
However, there is a particle-hole symmetry of the eigenstates of $H(\vec{k})$.
For the discussion here, we set $\epsilon(k)=0$ without loss of generality.
We postulate the operator
\newcommand{\Peh}{P_{eh}}
\begin{equation}
\label{Peh}
\Peh(\phi) = \left(\begin{array}{cccc}
                    &   k_+/k &  & \\
                -k_-/k   &  &   &   \\
                    &   & & -k_-/k \\
                     &  &  k_+/k &  
                  \end{array}\right)
\end{equation}
where $\phi = \arg(k_+) = \arg(k_x + i k_y)$ and $k = |\vec{k}|$.
We will show that $P_{eh}$ has the meaning of a particle-hole symmetry, in the sense of \eqref{PehProjectors}.
Obviously, $P_{eh}$ does not depend on $k$, and it has the properties $\Peh^\dagger = -\Peh = \Peh^{-1}$. 
Being interested in $H(\vec{k})$ as a function of the parameters $R$ and $T$, we denote this with an index.
We find the relation
\begin{equation}
\label{Pehsymmetry}
 -\Peh H_{RT} \Peh^\dagger = H_{-T k^{-2},-R k^2}.
\end{equation}
Thereby, we recognize that for $R = T =0$, the operator $P_{eh}$  gives the particle-hole symmetry $\{P_{eh}, \hat{H} \} = 0$.
Further, $[\Peh, \hat{h}] = 0$, as is easily checked using the representation \eqref{helop} in the main text. Therefore, if $\ket{\chi}$ is an helicity eigenstate, $\Peh \ket{\chi}$ is also an eigenstate of the same helicity.
Now we want to prove that if $\ket{\chi}$ is an energy eigenstate, $\Peh \ket{\chi}$ is also an energy eigenstate, which is not yet clear 
from \eqref{Pehsymmetry} for the case $R, T \ne 0$. Thus we can find the complete set of eigenvectors just starting with two states $\ket{\chi_\pm}$
having helicity $\pm 1$.

In the following, we use the notation $\ket{\chi_{n,\vec{k}}} = \ket{\chi_{n,\phi}}$ for energy eigenvectors, when dependency on $k$ is unimportant.
TRS relates states of $\vec{k}$ to states of $-\vec{k}$.
Since we are looking for a relation between states of the same $\vec{k}$, we use a combination of 
time reversal and a rotation by $\pi$. The rotational symmetry is $D_{-\phi} H(\phi) D_{\phi} = H(0)$.

Since energies at $k \ne 0$ are non-degenerate, we know that time reversal will map spinors onto themselves,
\footnote{
Here we consider just the spinor part of the state. The full state  $| \psi_{\vec{k}} > = |\vec{k}> |\chi_{n,\vec{k}}>$
 is of course mapped to it's 
Kramer's partner, which is attached at $|-\vec{k}>$ and thus is orthogonal to $|\vec{k}>$.
Note also that using the rotation to go from $\vec{k} \to -\vec{k}$ 
corresponds to the continuous change of the eigenvector 
on the circle $k = const.$, so we end up with a state of same energy. If we did a continuous change of the eigenvector by keeping $\phi = const.$ as in the previous section, we would end up with a state of different energy.
}
$Z \ket{\chi_{n,-\vec{k}}^*} = Z D_{\pi}^* \ket{\chi_{n,\vec{k}}^*} = e^{i \phi_n} \ket{\chi_{n,\vec{k}}}$. 
The state-dependent phase $e^{i \phi_n}$ is unimportant because it can be removed by re-definition.
Let us define $U := (Z D_{\pi}^*)^\dagger = i \sigma_x \tau_z$. Then we can replace the operation of complex conjugation by the unitary operation $U$,
 $\ket{\chi_n^*} = U \ket{\chi_n}$.
Since $[P_{eh}(0), U] = [i \sigma_z \tau_y, i \sigma_x \tau_z] = 0$, we exploit the rotational symmetry $D_{-\phi} P_{eh}(\phi) D_{\phi} = P_{eh}(0)$ 
to evaluate the commutator at $\phi=0$.

Our goal is to prove 
\begin{align}
\label{PehZero}
 \bra{\chi_{n,\phi}} P_{eh}(\phi) \ket{\chi_{n,\phi}} = 0.
\end{align}
Eigenstates of different energy, but same $\vec{k}$, must be orthogonal, and the SO terms remove degeneracy.
Since the eigenspaces of $h(\vec{k})$ are just 2-dimensional, finding a state of same helicity that is orthogonal, suffices to show that it is an eigenstate of energy.
Thus \eqref{PehZero} is equivalent to \eqref{PehProjectors}.
Since $P_{eh}^\dagger = -P_{eh}$, and $ \ket{\chi_{n,\phi}} = D_{\phi} \ket{\chi_{n,0}}$, we have 
\begin{align}
\nonumber & -\bra{\chi_{n,\phi}} P_{eh}(\phi) \ket{\chi_{n,\phi}} 
 = - \bra{\chi_{n,0}} P_{eh}(0) \ket{\chi_{n,0}} 
\\ \nonumber & = \bra{\chi_{n,0}} P_{eh}(0) \ket{\chi_{n,0}}^*
 =  \bra{\chi_{n,0}} U^\dagger P_{eh}^*(0) U \ket{\chi_{n,0}}
\\ & = \bra{\chi_{n,0}} U^\dagger P_{eh}(0) U \ket{\chi_{n,0}} = \bra{\chi_{n,0}} P_{eh}(0) \ket{\chi_{n,0}} = 0.
\end{align}

\subsection{Using $h(\vec{k})$ in the transport code}
\label{AppendixHelicityOp:transport}
For a state $\ket{\vec{k}}(a \ket{\chi_+(\vec{k})} + b \ket{\chi_-(\vec{k})})$ (which is not eigenstate of $\hat{H}$), 
we intend to define our spin transport by measuring $|a|^2 - |b|^2$. 
In transport calculations, one typically considers eigenstates of energy at the Fermi level. Then, the $\vec{k}$-vectors
in general will be different: 
\begin{align}
 \ket{\psi} = a \ket{\vec{k}_1} \ket{\chi_+(\vec{k}_1)} + b \ket{\vec{k}_2} \ket{\chi_-(\vec{k}_2)}
\end{align}
Here we include just two modes, because for typical parameters and a given direction of $\vec{k}$, $H(\vec{k})$ will have just two propagating modes
at the Fermi level, and we expect evanescent modes to have negligible effect on transport.
Because the Hamiltonian is rotationally invariant, the direction of propagation coincides with the direction of $\vec{k}$, and we assume 
$\vec{k}_1/k_1 = \vec{k}_2/k_2$.
The helicity operator just depends on the direction $\vec{k}/k$.
On the one hand, we have
\begin{align}
& \nonumber \bra{\psi} \hat{h} \ket{\psi} = \bra{\psi} \left(a \ket{\vec{k}_1} h(\vec{k}_1) \ket{\chi_+(\vec{k}_1)} + b \ket{\vec{k}_2} h(\vec{k}_2) \ket{\chi_-(\vec{k}_2)} \right)
\\&  = |a|^2 - |b|^2.
\end{align}
Instead of calculating this global expectation value, we may as well calculate the local expectation value $\bra{\chi} h(\vec{k}_i) \ket{\chi}$ (i=1 or 2) using just the 
4-component spinor  $\ket{\chi} := a \ket{\chi_+(\vec{k}_1)} + b \ket{\chi_-(\vec{k}_2)}$, to obtain the same result,
\begin{align}
& \nonumber \bra{\chi} h(\vec{k}_i) \ket{\chi}  
  =  |a|^2 - |b|^2  + b^* a \braket{\chi_-(\vec{k}_2)}{\chi_+ (\vec{k}_1)}
  \\&  - a^* b \braket{\chi_+(\vec{k}_1)}{\chi_- (\vec{k}_2)} 
  = |a|^2 - |b|^2.
\end{align}
Here we used the projector representation of $h(\vec{k})$ and made use of the fact that the expectation value must be real.
If we want to numerically evaluate the local expectation value of $h(\vec{k})$, we only need to 
choose a fixed direction of $\vec{k}$, and we can analyze the contribution of the modes for a wave 
going in this direction. 

\section{Obtaining bias-dependent observables from the S-matrix}
\label{AppendixObsSmat}
\newcommand{\ndens}{\hat{n}}

We have seen that the helicity, i.e. eigenvalue of $h(\vec{k})$ for a given direction of $\vec{k}$, corresponds to the 
components of the electron beam that will be split at the SO barrier.
Since the right lead is constructed in a way to guide the beam in the direction $k=k_x$, we consider the operator $h_x = h(k=k_x) 
= \sigma_y \tau_z$, 
for which the expectation value is called local helicity density, and the associated current is called helicity current. 
Although $[H(\vec{k}), h(\vec{k})] = 0$,  we have $[\hat{H}, h_x] \ne 0$ even with zero Rashba SO terms. 
Thus, the common practice to calculate spin currents, by introducing separate leads for the spin directions, does not work here. 

Instead, we show a method for the helicity current calculation, which combines transmission coefficients of the S-matrix 
with operator expectation values, evaluated for propagating states in the leads.
The advantage is that this method is even applicable with Rashba SO coupling in the leads. In that case, the local spin/helicity current oscillates as function of the 
position in the lead, but we will be interested in its average value only.

%
%
%
 

A lead is connected to a contact at one end and to the scattering region at the other end.
Thanks to the reflectionless property of the contacts, for a  lead $l$, the ingoing modes will be
populated with a Fermi distribution $f_0(E-\mu_l)$, while the outgoing modes will be populated by electrons
that originate from ingoing modes of other leads and which pass the scattering region.
  
Now we construct a density matrix $\rho$ for the ingoing states $\alpha,\beta$.
We take $\alpha$ as combined index $\alpha = (l_\alpha, k_\alpha, n_\alpha)$ of lead, momentum ($x$ component, i.e. along the direction of the lead) and mode index.
We use the short notations $\delta_{\alpha,\beta} = \delta_{l_\alpha,l_\beta} \delta(k_\alpha - k_\beta) \delta_{n_\alpha, n_\beta}$ and 
 $\int \rd\alpha = \sum_{l_\alpha} \int \rd k_\alpha \sum_{n_\alpha}$. $\epsilon_\alpha$ is the subband dispersion of lead $l_\alpha$ and $v_\alpha = \partial \epsilon_\alpha/\partial k_\alpha$ is the velocity of mode $n_\alpha$.
\begin{align}
 \rho = \int \rd\gamma \, f_0(\epsilon_\gamma - \mu_{l_\gamma})  \theta(v_\gamma)\ket{\gamma}\bra{\gamma} 
\end{align}
In order to obtain a density matrix $\rho'$ for the outgoing states, 
we propagate the states with the S-matrix like $\ket{\gamma} \to S \ket{\gamma}$
\begin{align}
 \rho'_{\alpha,\beta} =  \int \rd \gamma \, f_0(\epsilon_\gamma - \mu_{l_\gamma})
  \bra{\alpha} S \ket{\gamma}\bra{\gamma} S^\dagger \ket{\beta}
\end{align}
We have to specify how to evaluate S-matrix elements
\begin{align}
 \nonumber & \bra{\alpha} S \ket{\gamma} = \bra{k_\alpha, n_\alpha} S \ket{k_\gamma, n_\gamma} 
  \\ & = \sqrt{|v_\alpha v_\gamma|} \, \theta(v_\gamma) \theta(-v_\alpha) \delta(\epsilon_\alpha - \epsilon_\gamma) t_{l_\alpha n_\alpha, l_\gamma n_\gamma}
 \\&  \nonumber
  = \int d\mu' \frac{1}{\sqrt{|v_\alpha v_\gamma|}} 
  \delta(k_\alpha - k_\alpha(\mu') ) \delta(k_\gamma - k_\gamma(\mu'))  t_{l_\alpha n_\alpha, l_\gamma, n_\gamma}
\end{align}
where the current-normalized transmission amplitudes appear, i.e. the matrix of the $t_{l_\alpha n_\alpha, l_\gamma n_\gamma}$ is unitary,
and describes elastic scattering. The velocity factors ensure current conservation.
The $\theta$ factors select only ingoing modes in lead $l_\gamma$ and outgoing modes in lead $l_\alpha$, and 
the set of $k_\alpha(\mu')$ are the outgoing solutions of $\epsilon_\alpha(k) = \mu'$ for the set of subbands $n_\alpha$.

We are now prepared to evaluate a local observable $\hat{O}(x)$, $x$ lying in some lead $p$, relative to the lead's coordinate system.
We assume that all terms in $\hat{O}(x)$ contain a factor $P_x$ (projector on $x$).
We need the matrix elements in the basis of propagating (outgoing) states. They can be written in the form
\begin{align}
\nonumber & O_{\beta,\alpha} = \bra{\beta} \hat{O}(x) \ket{\alpha} 
\\& = \delta_{l_\alpha,p} \delta_{l_\beta,p} \bra{\chi_\beta} O(x, k_\beta,k_\alpha) \ket{\chi_\alpha} e^{i (k_\alpha - k_\beta) x}
\end{align}
with some matrix-valued function $O(x, k_\beta,k_\alpha)$ and the normalized transverse modes $\{ \ket{\chi_\alpha} \}$.
For the (helicity or spin) current operators, $O(x, k_\beta,k_\alpha)$ does not depend on $x$, 
and for the local (helicity or spin) density operators, it also does not depend on $k_\alpha$ and $k_\beta$.

Then, the expectation value of $\hat{O}(x)$ depends on in- and outgoing modes,
\begin{align}
 \langle \hat{O}(x) \rangle = \Tr [ (\rho + \rho') \hat{O}(x) ].
\end{align}
We leave equilibrium by introducing lead-dependent bias voltages $\mu_l = E_F + \delta \mu_l$.
If we calculate the response at lead $p$ due to some bias at lead $q\ne p$, $\rho$ will not contribute.
\begin{eqnarray}
\label{Oxdmu}
 \nonumber &&\frac{\delta \langle \hat{O}(x) \rangle }{\delta \mu_q}  
 =  \int \rd \alpha \int \rd \beta \, \frac{ \partial \rho'_{\alpha,\beta}}{\partial \mu_q}  O_{\beta,\alpha} 
\\ \nonumber &&
 =  \int \rd \alpha \int \rd \beta \int \rd \gamma \, \sqrt{| v_\alpha v_\beta |} |v_\gamma| t_{\alpha,\gamma} t_{\beta,\gamma}^* 
 \delta(\epsilon_\alpha - \epsilon_\gamma)
 \\  &&  \delta(\epsilon_\gamma - \epsilon_\beta)  \theta(-v_\alpha)\theta(-v_\beta) \theta(v_\gamma) \frac{\partial f_0(\epsilon_\gamma - \mu_\gamma)}{\partial \mu_p} O_{\beta,\alpha}
\end{eqnarray}
Let us assume zero temperature, so $\theta(v_\gamma) \frac{\partial f_0(\epsilon_\gamma - \mu_\gamma)}{\partial \mu_q} = \frac{1}{|v_\gamma|} \delta(k_\gamma - k_q^n) \delta_{l_\gamma,q}$, 
 where $k_q^n$ is the momentum of the ingoing mode $n$ in lead $q$ for energy $\mu_q$. 

In general, $O_{\beta,\alpha}$ will not be diagonal and therefore, it will also show oscillations in $x$.
But if $O(x,k_\beta,k_\alpha)$ is $x$-independent and we are interested in a mean value, we can still simplify \eqref{Oxdmu}
using
\begin{align}
\label{deltaaverage}
 \delta_{l_\alpha,l_\beta} \delta(\epsilon_\alpha - \epsilon_\beta) \int_0^L \rd x \, e^{-i (k_\alpha - k_\beta) x} \to \frac{L}{|v_\alpha|} \delta_{\alpha,\beta}
\end{align}
for large $L$. Put into words, if the lead and momentum indices are the same, the band indices must also be the same 
if energies are the same. Note that, if we have degeneracy of energy and momentum, this relation will not hold.
For the lead-averaged response of the expectation value (e.g. for the conductance), we find
\begin{eqnarray}
\nonumber  &&
\frac{1}{L} \int_0^L \rd x \frac{\delta \langle \hat{O}(x) \rangle }{\delta \mu_q} 
 =  \int \rd \alpha  \int \rd \gamma \, |v_\gamma| |t_{\alpha,\gamma}|^2 
 \\
 && \delta(\epsilon_\alpha - \epsilon_\gamma) \theta(v_\gamma) \frac{\partial f_0(\epsilon_\gamma - \mu_\gamma)}{\partial \mu_p}
  O_{\alpha,\alpha} \delta_{l_\alpha,p}
\\ 
&& \overset{T=0}{=}   \int \rd \alpha  \sum_{n_\gamma} \, |t_{\alpha,\gamma}|^2 \delta(\epsilon_\alpha - \epsilon_\gamma) 
  O_{\alpha,\alpha} \delta_{l_\alpha,p}
\\ 
\label{dOincoherent}
&& =   \sum_{n_\alpha}  \sum_{n_\gamma} \, |t_{\alpha,\gamma}|^2 
   \frac{1}{|v_\alpha|} O_{\alpha,\alpha}
\end{eqnarray}
where the solutions $k_\alpha(E_F)$ of lead $q$ should be entered whenever integration over $\alpha$ is no longer present.
The derivation shown here is mostly standard, apart from the averaging step \eqref{deltaaverage}. It is this step which 
keeps our result quite general and simple at the same time.
In the literature, most of the time $\hat{O}(x)$ is taken as the current operator.
Then, averaging is not necessary since $O_{\beta,\alpha}$ is already diagonal.

For the case of degeneracy in energy and momentum, the basis used to evaluate $O_{\alpha,\alpha}$
will matter, although numerical diagonalization will choose an arbitrary basis.
In particular, this applies to the situation without Rashba SO in the leads, where subbands are degenerate.
There are two ways to fix this problem: 
First, instead of \eqref{dOincoherent}, we can use
\begin{align}
\label{ObsfromSmat}
 \sum'_{m,n,l} t_{m,l} t_{n,l}^* \frac{1}{|v_n|} O_{n,m}
\end{align}
where we replaced the collective Greek indices by the Latin mode indices, being the only ones of interest here.
The sum is over all modes $m$,$n$ in lead $p$ and $l$ in lead $q$, both at the Fermi energy. Since we are interested only in average over the lead,
the summation over $m$,$n$ is restricted to pairs with $k_m = k_n$. Thus, also $v_n = v_m$.
If there are only two propagating modes, and if we do not include Rashba SO terms in the lead, i.e. when all modes are degenerate, 
formula \eqref{ObsfromSmat} will be also correct without the averaging over the $x$-coordinate (because the result is constant).

Alternatively, we may get rid of the degeneracy by adding a tiny perturbation which will fix the mode basis in the leads.
The form of the perturbation will depend on the operator $\hat{O}(x)$.
E.g, if we introduce a magnetic field $B_z$ as perturbation, this will cause spin precession about $z$ and therefore suppress the $\sigma_y$-polarization
even in the limit $B_z \to 0$, thus changing the physics.
In order to check that the small perturbation does not change the physics, we have
to prove that the general result \eqref{ObsfromSmat} reduces to \eqref{dOincoherent}
in the limit of the vanishing perturbation.

It turns out, in the 2-band model (2DEG with Rashba SO), when interested in $\sigma_y$-polarization or currents, we may use either a small magnetic field $B_y$ or a small Rashba SO coupling.
In the 4-band model, things are more complicated. But we find that again, a small Rashba SO term does the job and does not change the physics.

\section{Continuity eq. and vanishing average torque}
\label{AppendixContTorque}
We want to understand how the helicity current that we analyzed in the main text, is connected to the helicity polarization.
Therefore, in this section we derive a generalized continuity equation.
For a spin-$\frac{1}{2}$ system with Rashba SO,  the continuity equation including a source term reads \cite{ShiNiu06} 
\begin{align}
\label{contin2band}
 \frac{\partial S_l}{\partial t} + \nabla \cdot \vec{J}^{(l)} = \mathcal{T}_l
\end{align}
with the spin density $S_l = \psi^\dagger(r) \sigma_l \psi(r)$, spin current $\vec{J}^{(l)} = \Re \left(\psi^\dagger(r) \frac{1}{2i} \{ [r, H], \sigma
_l \} \psi(r) \right)$ and the 
spin source (torque) $\mathcal{T}_l = \Re \left( \psi^\dagger(r) \frac{1}{i} [\sigma_l, H] \psi(r) \right)$.

We generalize this result to the 4-band model \eqref{hgte2deg}, which is an effective model for the envelope function. 
Since the quadratic and cubic terms of $H_{SO}$ are unimportant (see main text), we consider only  the linear Rashba terms.
Our goal is to find an equation similar to \eqref{contin2band}, but the Pauli matrix $\sigma_l$ should be replaced by a general
$n \times n$ matrix $\Xi$ with constant entries (no operators).
To obtain helicity density, current, and torque terms, we may later choose $\Xi = h_x = \sigma_y \tau_z$. 
To find density, current and torque terms related to the spin-z polarization, we may choose $\Xi = \sigma_z \tau_0$.
We start with a standard derivation of current matrix elements, which is also appropriate for finding our $\Xi$-currents.

We consider the general $n$-band model 
\begin{align}
H = H_0 + V(\vop{r}) + U(\vop{r},\vop{k}) 
\end{align}
where $H_0 = \sum_i \epsilon_i(\hat{k}^2) \ket{i}\bra{i}$ is a diagonal matrix,
$V(\vop{r})$ is a general Hermitian $n \times n$ matrix representing a local potential, 
and $U$ is linear in $\hat{\vec{k}}$ and Hermitian, of the form
\begin{align}
 U(\vop{r},\vop{k}) = \sum_i \alpha_i(\vop{r}) \hat{k}_i \alpha_i(\vop{r})
\end{align}
with a vector $\boldsymbol\alpha(\vop{r})$ of $n \times n$ matrices, here with 2 components.
We note in passing that for another common symmetrization, $U_A(\vop{r},\vop{k}) = \sum_i \{\alpha_i(\vop{r}), \hat{k}_i\}$, 
the resulting continuity equation is of the same form.

For arbitrary wave functions $\psi_n$, $\psi_m$, we consider the overlap
\begin{align}
 w_{nm}(\vec{r}) = \psi_{n}^\dagger(\vec{r}) \psi_m(\vec{r})   = \braket{\psi_n}{\vec{r}}\braket{\vec{r}}{\psi_m}.
\end{align}
We apply the Schr\"odinger equation to get
\begin{align}
\label{deltwnm}
 \partial_t w_{nm}(\vec{r}) = 
\bra{\psi_n} \frac{1}{i} \Big[ \ket{\vec{r}}\bra{\vec{r}}, H  \Big] \ket{\psi_m}.
\end{align}
Here, terms with $V(\vop{r})$ cancel because they are local.
As usual, we exploit that $H_0$ consists of second derivatives $\nabla^2$.  Pulling out $\nabla$ 
and employing $\hat{\vec{k}} = \frac{1}{2i} [\hat{\vec{r}}, \hat{k}^2 ]$, we arrive at 
\begin{align}
 \nonumber &&\partial_t w_{nm}(\vec{r}) = 
 -\frac{1}{2} \nabla \bra{\psi_n} \Big\{ \ket{\vec{r}}\bra{\vec{r}},  \frac{1}{i} [\vop{r}, H_0] \Big\}  \ket{\psi_m} 
\\
&&+ \bra{\psi_n} \frac{1}{i} \Big[ \ket{\vec{r}}\bra{\vec{r}},  U \Big]  \ket{\psi_m}.
\end{align}
For the $U$ term, we use
\begin{eqnarray}
\nonumber && \bra{\psi_n} \frac{1}{i} \Big[ \ket{\vec{r}}\bra{\vec{r}},  U \Big]  \ket{\psi_m} 
 =  -\sum_j \partial_j  \braket{\psi_n}{\vec{r}} \alpha_j^2 \braket{\vec{r}}{\psi
_m} 
\\ && 
 = -\frac{1}{2}\nabla \bra{\psi_n} \Big\{ \ket{\vec{r}}\bra{\vec{r}}, \frac{1}{i} [ \vec{r}, U] \Big\} \ket{\psi_m}.
\end{eqnarray}
Since  $[\vop{r}, H_0] = [\vop{r}, H_0 + V(\vop{r})]$, we obtain the continuity equation $\partial_t w_{nm} = -\nabla \vec{j}_{nm}$ with
\newcommand{\Vop}{\mathcal{V}}
\begin{eqnarray}
&& \vec{j}_{nm}(\vec{r})  = \bra{\psi_n}  \frac{1}{2}\Big\{ \ket{\vec{r}}\bra{\vec{r}}, \boldsymbol\Vop \Big\} \ket{\psi_m},  
\end{eqnarray}
with $\Vop_j = \frac{\partial H}{\partial k_j} = \frac{1}{i} [ \hat{r}_j, H]$.
The diagonals are real and give the well-known current expression
\begin{align}
 \vec{j}_{nn}(\vec{r})  = \Re \left( \psi_n^\dagger (\vec{r}) \frac{1}{i} [\vop{r}, H] \psi_n(\vec{r}) \right).
\end{align}

Based on this calculation, is it not difficult to find  our generalized continuity equation for the current of $\Xi$ (e.g. helicity),
\begin{align}
 \frac{\partial S_{\Xi}}{\partial t} + \nabla \cdot \vec{J}^{(\Xi)} = \mathcal{T}_{\Xi}
\end{align}
with the $\Xi$-density $S_{\Xi} = \psi^\dagger(\vec{r}) \Xi \psi(\vec{r})$, 
$\Xi$-current $\vec{J}^{(\Xi)} = \Re \left(\psi^\dagger(\vec{r}) \frac{1}{2i} \{ [\vop{r}, H], \Xi  \} \psi(\vec{r})\right)$ 
and the $\Xi$-source (torque) $\mathcal{T}_{\Xi}  = \Re \left(\psi^\dagger(\vec{r}) \frac{1}{i} [\Xi, H] \psi(\vec{r})\right)$.

We define the projector on coordinate $\vec{r}$, 
$P_{\vec{r}} = \ket{\vec{r}}\bra{\vec{r}} $. With the Schr\"odinger equation and it's Hermitian conjugate, we obtain, similar to \eqref{deltwnm}
\begin{align}
 \partial_t \braket{\psi_n}{\vec{r}} \Xi  \braket{\vec{r}}{\psi_m} 
= \frac{1}{i} \bra{\psi_n} [P_{\vec{r}} \Xi, H] \ket{\psi_m}.
\end{align}
The diagonal element ($m=n$) gives $\frac{\partial S_{\Xi}}{\partial t}$.
Since $[P_{\vec{r}}, \Xi] = 0$, we have 
\begin{align}
 \label{Prhxtime}
 \frac{1}{i} [ P_{\vec{r}} \Xi, H] 
 = \frac{1}{2i} \{ [P_{\vec{r}}, H], \Xi \}  + \frac{1}{2i} \{P_{\vec{r}}, [\Xi, H]\}
\end{align}
The diagonal matrix element of the last term gives the $\Xi$-torque
$\mathcal{T}_{\Xi} = \frac{1}{2i} \bra{\psi_n} \{P_{\vec{r}}, [\Xi, H] \} \ket{\psi_n}$.

For the first summand in \eqref{Prhxtime}, we note that 
the steps following \eqref{deltwnm} stay valid if we substitute $\ket{\psi_m} \to \Xi \ket{\psi_m}$ or 
$\bra{\psi_n} \to \bra{\psi_n} \Xi$, and therefore, 
\begin{align}
 &&\bra{\psi_n} \frac{1}{2} \{ [P_{\vec{r}}, H], \Xi \} \ket{\psi_m} = - \nabla \cdot \vec{J}^{(\Xi)}_{nm} 
\end{align}
with 
\begin{align}
 \vec{J}^{(\Xi)}_{nm}  = \frac{1}{4}  \bra{\psi_n} \{ \{P_{\vec{r}}, \boldsymbol{\mathcal{V}} \}, \Xi \} \ket{\psi_m}
 = \frac{1}{4} \bra{\psi_n} \{ \{\Xi, \boldsymbol{\mathcal{V}} \}, P_{\vec{r}} \} \ket{\psi_m}
\end{align}
The matrix element with $n=m$ again gives the $\Xi$-current $\vec{J}^{(\Xi)}$, which is real. 
To summarize, the derivation holds as long as on operator $\Xi$ fulfills the conditions $[\Xi, \vop{r}] = 0$ and $[\Xi, \vop{k}] = 0$, since otherwise, 
the derivative would also act on $\Xi$.

Finally, we note that the average $\Xi$-torque, when we evaluate it with an energy eigenstate and integrate over both coordinates of a lead, vanishes. 
For this, we simply need to use  $\int dx \int dy\,  P_{\vec{r}} = 1$.
Then, 
$ \int dx \int dy \, \mathcal{T}_{\Xi} = \bra{\psi} \frac{1}{i} [\Xi, H] \ket{\psi} = 0$ since $H \ket{\psi} = E \ket{\psi}$.
Applying this to $\Xi = h_x$, we see that the average helicity current is a conserved quantity.

\section{Wave matching for lattice model}
\label{AppendixWaveMat}
We consider an infinite N-SO or SO-SO interface. 
By employing an approximation of the analytical model by a lattice, we prevent issues with symmetrization \cite{Winkler93} and avoid problems with 
unphysical spurious solutions \cite{Schuurmans85}.
We use the Green's function formalism as developed e.g. in \cite{WimmerPHD} for a tight-binding model with only nearest neighbor hoppings. 
Since the formalism is the same as is used also in the finite geometry with attached leads, 
we may call the left and right sides of the interface two ``leads''.
Due to $k_y$-conservation, we only need to solve a 1D chain problem with 
$k_y$ as parameter, while $\hat{k}_x$ will be discretized on a lattice with lattice constant $a$.
Due to translational invariance by shift of $a$, solutions $\psi(x)$ will be plane waves 
with $k_x$ in the first Brillouin zone $[-\pi/a, \pi/a]$. 
However, the 4-band model Hamiltonian \eqref{hgte2deg} contains up to 3rd powers of $\hat{k}_x$, which when discretized, lead 
to next-nearest neighbor hopping elements. Since the formalism is formulated in nearest-neighbor 
coupling matrices only, we need to use an enlarged unit cell containing 2 lattice sites (we identify them as sublattice A,B). So for 
the moment, we only make use of translational invariance by $2 a$. The ansatz $\psi(x) = e^{i k_x x} \chi(k_x)$
with a spinor $\chi(k_x)$ of 8 components, 
leads to the effective Schr\"odinger equation
\begin{equation}
\label{Hk}
 H(k_x) \chi(k_x)  = (H_0 + H_1 e^{2 i k_x a} + H_{-1} e^{-2 i k a}) \chi(k_x) = E \chi(k_x)
\end{equation}
with $H_{-1} = H_1^\dagger$. $H_0$, $H_1$ and $H_{-1}$ are 8x8 matrices describing the 
Hamiltonian of an isolated enlarged unit cell and the couplings to the right/left cells.

The band structure of $H(k_x)$ is formally obtained from the band structure of the primitive (i.e. single site) 
unit cell lattice problem by reducing the Brillouin zone to $[-\pi/(2a), \pi/(2a)]$ in the manner
of shifting $k_x$-values  by $\pi/a$ if necessary.
 So we have twice the number of bands in order to compensate for just half of the 
original Brillouin zone. 
We call the bands that are obtained by shifting (they originally have $|\Re(k_x)|>\pi/(2a)$) antibonding, 
and the other bonding.
We denote the components $[\chi]_{s,\alpha}$ of $\chi$ with an index $s=0,1=A,B$ for the sublattice 
and $\alpha=1,...,4$ for the band basis in which $\hat{H}$ is written.
For the bonding states we have $[\chi]_{A,\alpha}(k_x) = e^{i k_x a} [\chi]_{B,\alpha}(k_x)$ and 
for the antibonding states, $[\chi]_{A,\alpha}(k_x) = -e^{i k_x a} [\chi]_{B,\alpha}(k_x)$.
We can use these relations to find the value of $k_x$ in the primitive unit cell model.

Since we have to resort to hoppings by $2 a$ anyway, and we finally use the lattice model 
as an approximation of the analytical $\vec{k}$-diagonal model, we also use hoppings up to $2a$ to 
find the discretization of $\hat{k}_x$ and $\hat{k}_x^2$.
The discretization of $\hat{k}_x^2$ is found by fitting the parameters $c_j$ for hopping by $j a$ 
in the most general symmetric dispersion $E(k_x) = c_0 + c_1 \cos(k_x) + c_2 \cos(2 k_x)$. 
We find $c_0 = 5/2$, $c_1 = -8/3$ and $c_2 = 1/6$. Likewise, the representation of $\hat{k}_x$ 
is found by fitting $E(k_x) = d_1 \sin(k_x) + d_2 \sin(2 k_x)$, and we find $d_1 = 4/3$ and 
$d_2 = -1/6$.
This does not make the calculations more difficult, but gives a much better approximation to the 
continuum model.

We numerically solve \eqref{Hk} for a fixed energy $E_F$ and find the modes $\chi_{n} = \chi(k_n)$. 
The modes are classified in propagating with $|\lambda| = 1$ and evanescent modes with $|\lambda| \ne 1$,
and $\lambda = e^{2 i k_x a}$.
Further they are classified in right-going which is right-decaying ($|\lambda| < 1$) or right-moving ($|\lambda = 1$ and velocity $v >0$), and left-going.
The velocity of a normalized propagating mode $\chi(k_x)$ is 
obtained by $v = \chi^\dagger(k_x) \frac{\partial H(k_x)}{\partial k_x} \chi(k_x)$.
This relation still holds even if the expression 
$\psi^\dagger(\vec{r}) \frac{\partial H(k_x)}{\partial k_x} \psi(\vec{r})$ cannot be longer 
interpreted as local current density. This is the case for Hamiltonians containing powers of $\hat{k}_x$ higher
than two \cite{Li07}.
We calculate the helicity of a mode by putting the value of $k_x$ into the analytical expression of $h(\vec{k})$ 
and evaluating the expectation value just for sublattice A ($[\chi]_A$ is a 4-component vector): 
\begin{align}
 \langle \hat{h} \rangle_{\chi(\vec{k})} = \frac{[\chi]_A^\dagger h(\vec{k}) [\chi]_A}{[\chi]_A^\dagger [\chi]_A}
\end{align}
Here it is essential to put in the $k_x$-value obtained for the primitive lattice, by identifying
the bonding/antibonding character of $\chi$.
Because the rotational symmetry is broken, we do not expect to have perfect values $\pm 1$ for the helicity, but it turns out that the approximation to the analytical 
model works quite well - the helicity expectation value deviates from $\pm 1$ by less than $10^{-6}$.

Next, we use the eigenmodes to calculate the self-energies $\Sigma_R$, $\Sigma_L$ of the right and 
left lead. Of course, we have to calculate the modes separately for left and right lead if the parameters
of the Hamiltonian are different. But here we just use modes of the right lead to present formulas for both.
With notation of \cite{WimmerPHD}, ``$>$'' stands for outgoing and ``$<$'' for ingoing states, so at the right lead,
``$>$'' stands for left-going and ``$<$'' is right-going. 
The $16$ modes are sorted into two matrices, $U_> = (\chi_{1,>},...,\chi_{8,>})$ and $U_< = (\chi_{1,<},...,\chi_{8,<})$.
The corresponding eigenvalues $\lambda_n$ define matrices $\Lambda_> = \mathrm{diag}(\lambda_{1,>},...,\lambda_{8,>})$ 
and $\Lambda_< = \mathrm{diag}(\lambda_{1,<},...,\lambda_{8,<})$. Then we have \cite{WimmerPHD}
\begin{align}
 \nonumber
 \Sigma_R = H_1 U_< \Lambda_< U_<^{-1}, \quad \Sigma_L = H_{-1} U_> \Lambda_>^{-1} U_>^{-1}.
\end{align}
We can also obtain $\Sigma_L$ from $\Sigma_R$ by a rotation about $\pi$. For that, we have to combine 
rotations acting on band and sublattice space and $k_y \to -k_y$.
The corresponding $\Gamma$ matrices are $\Gamma_R = i(\Sigma_R - \Sigma_R^\dagger)$ and
$\Gamma_L = i(\Sigma_L - \Sigma_L^\dagger)$. It is very important to use right-decaying states for the 
right self energy and left-decaying states for the left self energy (evanescent states constitute 
the Hermitian part of  $\Sigma_{L,R}$). The $\Gamma$ matrices, on the 
other hand, project only on the propagating states, and e.g. $\Gamma_R$ may be rewritten with 
right-propagating states or left-propagating states. 

Finally we need the retarded Green's function $G^R = (E_F - H_0 - \Sigma_R - \Sigma_L)^{-1}$ 
of a single unit cell with the leads attached, modeled by the self-energies. 
Here we have the possibility to put in different Rashba SO values for $H_0$ in order to control the smoothness of the interface, 
e.g. we may use an average of the left and right lead's SO parameters for a smoothed interface. It turns out that cross-helicity transmissions (see main text)
decrease upon making the interface smoother.

The full scattering matrix can be obtained with the generalized Fisher-Lee relations \cite{WimmerPHD}
for transmission and reflection coefficients. For the left-to-right transmission amplitudes, right-going modes 
in the left lead are matched with right-going modes of the right lead. For the left-to-left reflection 
amplitudes, we match right-going modes with left-going modes in the left lead:
\begin{align}
 \label{tFisherLee}
 &t_{L;m,n} = \frac{i}{\sqrt{|v_{R;m,<} v_{L;n,>}|}} \chi_{R;m,<}^\dagger \, \Gamma_R \, G^R \, \Gamma_L \, \chi_{L;n,>}
 \\&
 \label{rFisherLee}
 r_{L;m,n} = \frac{1}{\sqrt{|v_{L;m,<} v_{L;n,>}|}} 
 \chi_{L;m,<}^\dagger  \left( i \Gamma_L \, G^R \, \Gamma_L  - \Gamma_L  \right) \chi_{L;n,>}
\end{align}

\bibliographystyle{apsrev}
\bibliography{notes}

\begin{thebibliography}{41}
\expandafter\ifx\csname natexlab\endcsname\relax\def\natexlab#1{#1}\fi
\expandafter\ifx\csname bibnamefont\endcsname\relax
  \def\bibnamefont#1{#1}\fi
\expandafter\ifx\csname bibfnamefont\endcsname\relax
  \def\bibfnamefont#1{#1}\fi
\expandafter\ifx\csname citenamefont\endcsname\relax
  \def\citenamefont#1{#1}\fi
\expandafter\ifx\csname url\endcsname\relax
  \def\url#1{\texttt{#1}}\fi
\expandafter\ifx\csname urlprefix\endcsname\relax\def\urlprefix{URL }\fi
\providecommand{\bibinfo}[2]{#2}
\providecommand{\eprint}[2][]{\url{#2}}

\bibitem[{\citenamefont{Wolf et~al.}(2001)\citenamefont{Wolf, Awschalom,
  Buhrman, Daughton, von Molnár, Roukes, Chtchelkanova, and Treger}}]{Wolf01}
\bibinfo{author}{\bibfnamefont{S.~A.} \bibnamefont{Wolf}},
  \bibinfo{author}{\bibfnamefont{D.~D.} \bibnamefont{Awschalom}},
  \bibinfo{author}{\bibfnamefont{R.~A.} \bibnamefont{Buhrman}},
  \bibinfo{author}{\bibfnamefont{J.~M.} \bibnamefont{Daughton}},
  \bibinfo{author}{\bibfnamefont{S.}~\bibnamefont{von Molnár}},
  \bibinfo{author}{\bibfnamefont{M.~L.} \bibnamefont{Roukes}},
  \bibinfo{author}{\bibfnamefont{A.~Y.} \bibnamefont{Chtchelkanova}},
  \bibnamefont{and} \bibinfo{author}{\bibfnamefont{D.~M.}
  \bibnamefont{Treger}}, \bibinfo{journal}{Science}
  \textbf{\bibinfo{volume}{294}}, \bibinfo{pages}{1488} (\bibinfo{year}{2001}),
  \urlprefix\url{http://www.sciencemag.org/content/294/5546/1488.abstract}.

\bibitem[{\citenamefont{Schmidt et~al.}(2000)\citenamefont{Schmidt, Ferrand,
  Molenkamp, Filip, and van Wees}}]{Schmidt2000}
\bibinfo{author}{\bibfnamefont{G.}~\bibnamefont{Schmidt}},
  \bibinfo{author}{\bibfnamefont{D.}~\bibnamefont{Ferrand}},
  \bibinfo{author}{\bibfnamefont{L.~W.} \bibnamefont{Molenkamp}},
  \bibinfo{author}{\bibfnamefont{A.~T.} \bibnamefont{Filip}}, \bibnamefont{and}
  \bibinfo{author}{\bibfnamefont{B.~J.} \bibnamefont{van Wees}},
  \bibinfo{journal}{Phys. Rev. B} \textbf{\bibinfo{volume}{62}},
  \bibinfo{pages}{R4790} (\bibinfo{year}{2000}),
  \urlprefix\url{http://link.aps.org/doi/10.1103/PhysRevB.62.R4790}.

\bibitem[{\citenamefont{Alvarado and Renaud}(1992)}]{Alvarado92}
\bibinfo{author}{\bibfnamefont{S.~F.} \bibnamefont{Alvarado}} \bibnamefont{and}
  \bibinfo{author}{\bibfnamefont{P.}~\bibnamefont{Renaud}},
  \bibinfo{journal}{Phys. Rev. Lett.} \textbf{\bibinfo{volume}{68}},
  \bibinfo{pages}{1387} (\bibinfo{year}{1992}),
  \urlprefix\url{http://link.aps.org/doi/10.1103/PhysRevLett.68.1387}.

\bibitem[{\citenamefont{Ohno et~al.}(1999)\citenamefont{Ohno, Young, Beschoten,
  Matsukura, Ohno, and Awschalom}}]{Ohno99}
\bibinfo{author}{\bibfnamefont{Y.}~\bibnamefont{Ohno}},
  \bibinfo{author}{\bibfnamefont{D.~K.} \bibnamefont{Young}},
  \bibinfo{author}{\bibfnamefont{B.}~\bibnamefont{Beschoten}},
  \bibinfo{author}{\bibfnamefont{F.}~\bibnamefont{Matsukura}},
  \bibinfo{author}{\bibfnamefont{H.}~\bibnamefont{Ohno}}, \bibnamefont{and}
  \bibinfo{author}{\bibfnamefont{D.~D.} \bibnamefont{Awschalom}},
  \bibinfo{journal}{Nature} \textbf{\bibinfo{volume}{402}},
  \bibinfo{pages}{790} (\bibinfo{year}{1999}).

\bibitem[{\citenamefont{Dyakonov and Perel}(1971)}]{Dyakonov2}
\bibinfo{author}{\bibfnamefont{M.~I.} \bibnamefont{Dyakonov}} \bibnamefont{and}
  \bibinfo{author}{\bibfnamefont{V.~I.} \bibnamefont{Perel}},
  \bibinfo{journal}{Phys. Lett.} \textbf{\bibinfo{volume}{A35}},
  \bibinfo{pages}{459} (\bibinfo{year}{1971}).

\bibitem[{\citenamefont{Hirsch}(1999)}]{Hirsch99}
\bibinfo{author}{\bibfnamefont{J.~E.} \bibnamefont{Hirsch}},
  \bibinfo{journal}{Phys. Rev. Lett.} \textbf{\bibinfo{volume}{83}},
  \bibinfo{pages}{1834} (\bibinfo{year}{1999}).

\bibitem[{\citenamefont{Murakami et~al.}(2003)\citenamefont{Murakami, Nagaosa,
  and Zhang}}]{Murakami03}
\bibinfo{author}{\bibfnamefont{S.}~\bibnamefont{Murakami}},
  \bibinfo{author}{\bibfnamefont{N.}~\bibnamefont{Nagaosa}}, \bibnamefont{and}
  \bibinfo{author}{\bibfnamefont{S.~C.} \bibnamefont{Zhang}},
  \bibinfo{journal}{Science} \textbf{\bibinfo{volume}{301}},
  \bibinfo{pages}{1348} (\bibinfo{year}{2003}),
  \urlprefix\url{http://www.sciencemag.org/cgi/content/abstract/301/5638/1348}.

\bibitem[{\citenamefont{Sinova et~al.}(2004)\citenamefont{Sinova, Culcer, Niu,
  Sinitsyn, Jungwirth, and MacDonald}}]{Sinova04}
\bibinfo{author}{\bibfnamefont{J.}~\bibnamefont{Sinova}},
  \bibinfo{author}{\bibfnamefont{D.}~\bibnamefont{Culcer}},
  \bibinfo{author}{\bibfnamefont{Q.}~\bibnamefont{Niu}},
  \bibinfo{author}{\bibfnamefont{N.~A.} \bibnamefont{Sinitsyn}},
  \bibinfo{author}{\bibfnamefont{T.}~\bibnamefont{Jungwirth}},
  \bibnamefont{and} \bibinfo{author}{\bibfnamefont{A.~H.}
  \bibnamefont{MacDonald}}, \bibinfo{journal}{Phys. Rev. Lett.}
  \textbf{\bibinfo{volume}{92}}, \bibinfo{pages}{126603}
  (\bibinfo{year}{2004}).

\bibitem[{\citenamefont{Kato et~al.}(2004)\citenamefont{Kato, Myers, Gossard,
  and Awschalom}}]{Kato}
\bibinfo{author}{\bibfnamefont{Y.}~\bibnamefont{Kato}},
  \bibinfo{author}{\bibfnamefont{R.~C.} \bibnamefont{Myers}},
  \bibinfo{author}{\bibfnamefont{A.~C.} \bibnamefont{Gossard}},
  \bibnamefont{and} \bibinfo{author}{\bibfnamefont{D.~D.}
  \bibnamefont{Awschalom}}, \bibinfo{journal}{Science}
  \textbf{\bibinfo{volume}{306}}, \bibinfo{pages}{1910} (\bibinfo{year}{2004}).

\bibitem[{\citenamefont{Wunderlich et~al.}(2005)\citenamefont{Wunderlich,
  Kaestner, Sinova, and Jungwirth}}]{Wunderlich}
\bibinfo{author}{\bibfnamefont{J.}~\bibnamefont{Wunderlich}},
  \bibinfo{author}{\bibfnamefont{B.}~\bibnamefont{Kaestner}},
  \bibinfo{author}{\bibfnamefont{J.}~\bibnamefont{Sinova}}, \bibnamefont{and}
  \bibinfo{author}{\bibfnamefont{T.}~\bibnamefont{Jungwirth}},
  \bibinfo{journal}{Phys. Rev. Lett.} \textbf{\bibinfo{volume}{94}},
  \bibinfo{pages}{047204} (\bibinfo{year}{2005}).

\bibitem[{\citenamefont{Valenzuela and Tinkham}(2006)}]{Tinkham06}
\bibinfo{author}{\bibfnamefont{S.~O.} \bibnamefont{Valenzuela}}
  \bibnamefont{and} \bibinfo{author}{\bibfnamefont{M.}~\bibnamefont{Tinkham}},
  \bibinfo{journal}{Nature} \textbf{\bibinfo{volume}{442}},
  \bibinfo{pages}{176} (\bibinfo{year}{2006}),
  \urlprefix\url{http://dx.doi.org/10.1038/nature04937}.

\bibitem[{\citenamefont{Br{\"u}ne et~al.}(2010)\citenamefont{Br{\"u}ne, Roth,
  Novik, Koenig, Buhmann, Hankiewicz, Hanke, Sinova, and Molenkamp}}]{Bruene10}
\bibinfo{author}{\bibfnamefont{C.}~\bibnamefont{Br{\"u}ne}},
  \bibinfo{author}{\bibfnamefont{A.}~\bibnamefont{Roth}},
  \bibinfo{author}{\bibfnamefont{E.~G.} \bibnamefont{Novik}},
  \bibinfo{author}{\bibfnamefont{M.}~\bibnamefont{Koenig}},
  \bibinfo{author}{\bibfnamefont{H.}~\bibnamefont{Buhmann}},
  \bibinfo{author}{\bibfnamefont{E.~M.} \bibnamefont{Hankiewicz}},
  \bibinfo{author}{\bibfnamefont{W.}~\bibnamefont{Hanke}},
  \bibinfo{author}{\bibfnamefont{J.}~\bibnamefont{Sinova}}, \bibnamefont{and}
  \bibinfo{author}{\bibfnamefont{L.~W.} \bibnamefont{Molenkamp}},
  \bibinfo{journal}{Nature Physics} \textbf{\bibinfo{volume}{6}},
  \bibinfo{pages}{448} (\bibinfo{year}{2010}).

\bibitem[{\citenamefont{Tse and Das~Sarma}(2006)}]{TseDSarma06}
\bibinfo{author}{\bibfnamefont{W.-K.} \bibnamefont{Tse}} \bibnamefont{and}
  \bibinfo{author}{\bibfnamefont{S.}~\bibnamefont{Das~Sarma}},
  \bibinfo{journal}{Phys. Rev. B} \textbf{\bibinfo{volume}{74}},
  \bibinfo{pages}{245309} (\bibinfo{year}{2006}),
  \urlprefix\url{http://link.aps.org/doi/10.1103/PhysRevB.74.245309}.

\bibitem[{\citenamefont{Hankiewicz and Vignale}(2008)}]{Hankiewicz08}
\bibinfo{author}{\bibfnamefont{E.~M.} \bibnamefont{Hankiewicz}}
  \bibnamefont{and} \bibinfo{author}{\bibfnamefont{G.}~\bibnamefont{Vignale}},
  \bibinfo{journal}{Physical Review Letters} \textbf{\bibinfo{volume}{100}},
  \bibinfo{eid}{026602} (pages~\bibinfo{numpages}{4}) (\bibinfo{year}{2008}),
  \urlprefix\url{http://link.aps.org/abstract/PRL/v100/e026602}.

\bibitem[{\citenamefont{Hankiewicz and Vignale}(2009)}]{HankiewiczVignale09}
\bibinfo{author}{\bibfnamefont{E.~M.} \bibnamefont{Hankiewicz}}
  \bibnamefont{and} \bibinfo{author}{\bibfnamefont{G.}~\bibnamefont{Vignale}},
  \bibinfo{journal}{Journal of Physics: Condensed Matter}
  \textbf{\bibinfo{volume}{21}}, \bibinfo{pages}{253202}
  (\bibinfo{year}{2009}),
  \urlprefix\url{http://stacks.iop.org/0953-8984/21/i=25/a=253202}.

\bibitem[{\citenamefont{Bi et~al.}(2013)\citenamefont{Bi, He, Hankiewicz,
  Winkler, Vignale, and Culcer}}]{BiCulcer13}
\bibinfo{author}{\bibfnamefont{X.}~\bibnamefont{Bi}},
  \bibinfo{author}{\bibfnamefont{P.}~\bibnamefont{He}},
  \bibinfo{author}{\bibfnamefont{E.~M.} \bibnamefont{Hankiewicz}},
  \bibinfo{author}{\bibfnamefont{R.}~\bibnamefont{Winkler}},
  \bibinfo{author}{\bibfnamefont{G.}~\bibnamefont{Vignale}}, \bibnamefont{and}
  \bibinfo{author}{\bibfnamefont{D.}~\bibnamefont{Culcer}},
  \bibinfo{journal}{Phys. Rev. B} \textbf{\bibinfo{volume}{88}},
  \bibinfo{pages}{035316} (\bibinfo{year}{2013}),
  \urlprefix\url{http://link.aps.org/doi/10.1103/PhysRevB.88.035316}.

\bibitem[{\citenamefont{Governale et~al.}(2003)\citenamefont{Governale, Taddei,
  and Fazio}}]{Governale03}
\bibinfo{author}{\bibfnamefont{M.}~\bibnamefont{Governale}},
  \bibinfo{author}{\bibfnamefont{F.}~\bibnamefont{Taddei}}, \bibnamefont{and}
  \bibinfo{author}{\bibfnamefont{R.}~\bibnamefont{Fazio}},
  \bibinfo{journal}{Phys. Rev. B} \textbf{\bibinfo{volume}{68}},
  \bibinfo{pages}{155324} (\bibinfo{year}{2003}),
  \urlprefix\url{http://link.aps.org/doi/10.1103/PhysRevB.68.155324}.

\bibitem[{\citenamefont{Brosco et~al.}(2010)\citenamefont{Brosco, Jerger,
  San-Jos\'e, Zarand, Shnirman, and Sch\"on}}]{Brosco10}
\bibinfo{author}{\bibfnamefont{V.}~\bibnamefont{Brosco}},
  \bibinfo{author}{\bibfnamefont{M.}~\bibnamefont{Jerger}},
  \bibinfo{author}{\bibfnamefont{P.}~\bibnamefont{San-Jos\'e}},
  \bibinfo{author}{\bibfnamefont{G.}~\bibnamefont{Zarand}},
  \bibinfo{author}{\bibfnamefont{A.}~\bibnamefont{Shnirman}}, \bibnamefont{and}
  \bibinfo{author}{\bibfnamefont{G.}~\bibnamefont{Sch\"on}},
  \bibinfo{journal}{Phys. Rev. B} \textbf{\bibinfo{volume}{82}},
  \bibinfo{pages}{041309} (\bibinfo{year}{2010}),
  \urlprefix\url{http://link.aps.org/doi/10.1103/PhysRevB.82.041309}.

\bibitem[{\citenamefont{Watson et~al.}(2003)\citenamefont{Watson, Potok,
  Marcus, and Umansky}}]{Watson03}
\bibinfo{author}{\bibfnamefont{S.~K.} \bibnamefont{Watson}},
  \bibinfo{author}{\bibfnamefont{R.~M.} \bibnamefont{Potok}},
  \bibinfo{author}{\bibfnamefont{C.~M.} \bibnamefont{Marcus}},
  \bibnamefont{and} \bibinfo{author}{\bibfnamefont{V.}~\bibnamefont{Umansky}},
  \bibinfo{journal}{Phys. Rev. Lett.} \textbf{\bibinfo{volume}{91}},
  \bibinfo{pages}{258301} (\bibinfo{year}{2003}),
  \urlprefix\url{http://link.aps.org/doi/10.1103/PhysRevLett.91.258301}.

\bibitem[{\citenamefont{Khodas et~al.}(2004)\citenamefont{Khodas, Shekhter, and
  Finkel'stein}}]{Khodas04}
\bibinfo{author}{\bibfnamefont{M.}~\bibnamefont{Khodas}},
  \bibinfo{author}{\bibfnamefont{A.}~\bibnamefont{Shekhter}}, \bibnamefont{and}
  \bibinfo{author}{\bibfnamefont{A.~M.} \bibnamefont{Finkel'stein}},
  \bibinfo{journal}{Phys. Rev. Lett.} \textbf{\bibinfo{volume}{92}},
  \bibinfo{pages}{086602} (\bibinfo{year}{2004}),
  \urlprefix\url{http://link.aps.org/doi/10.1103/PhysRevLett.92.086602}.

\bibitem[{\citenamefont{Bercioux and De~Martino}(2010)}]{Bercioux10}
\bibinfo{author}{\bibfnamefont{D.}~\bibnamefont{Bercioux}} \bibnamefont{and}
  \bibinfo{author}{\bibfnamefont{A.}~\bibnamefont{De~Martino}},
  \bibinfo{journal}{Phys. Rev. B} \textbf{\bibinfo{volume}{81}},
  \bibinfo{pages}{165410} (\bibinfo{year}{2010}),
  \urlprefix\url{http://link.aps.org/doi/10.1103/PhysRevB.81.165410}.

\bibitem[{\citenamefont{Bercioux et~al.}(2012)\citenamefont{Bercioux, Urban,
  Romeo, and Citro}}]{Bercioux12}
\bibinfo{author}{\bibfnamefont{D.}~\bibnamefont{Bercioux}},
  \bibinfo{author}{\bibfnamefont{D.~F.} \bibnamefont{Urban}},
  \bibinfo{author}{\bibfnamefont{F.}~\bibnamefont{Romeo}}, \bibnamefont{and}
  \bibinfo{author}{\bibfnamefont{R.}~\bibnamefont{Citro}},
  \bibinfo{journal}{Applied Physics Letters} \textbf{\bibinfo{volume}{101}},
  \bibinfo{eid}{122405} (pages~\bibinfo{numpages}{4}) (\bibinfo{year}{2012}),
  \urlprefix\url{http://link.aip.org/link/?APL/101/122405/1}.

\bibitem[{\citenamefont{Asmar and Ulloa}(2013)}]{Asmar13}
\bibinfo{author}{\bibfnamefont{M.~M.} \bibnamefont{Asmar}} \bibnamefont{and}
  \bibinfo{author}{\bibfnamefont{S.~E.} \bibnamefont{Ulloa}},
  \bibinfo{journal}{Phys. Rev. B} \textbf{\bibinfo{volume}{87}},
  \bibinfo{pages}{075420} (\bibinfo{year}{2013}),
  \urlprefix\url{http://link.aps.org/doi/10.1103/PhysRevB.87.075420}.

\bibitem[{\citenamefont{Kane and Mele}(2005)}]{Kane05QSHE}
\bibinfo{author}{\bibfnamefont{C.~L.} \bibnamefont{Kane}} \bibnamefont{and}
  \bibinfo{author}{\bibfnamefont{E.~J.} \bibnamefont{Mele}},
  \bibinfo{journal}{Phys. Rev. Lett.} \textbf{\bibinfo{volume}{95}},
  \bibinfo{pages}{226801} (\bibinfo{year}{2005}),
  \urlprefix\url{http://link.aps.org/doi/10.1103/PhysRevLett.95.226801}.

\bibitem[{\citenamefont{Bernevig et~al.}(2006)\citenamefont{Bernevig, Hughes,
  and Zhang}}]{Bernevig06}
\bibinfo{author}{\bibfnamefont{B.~A.} \bibnamefont{Bernevig}},
  \bibinfo{author}{\bibfnamefont{T.~L.} \bibnamefont{Hughes}},
  \bibnamefont{and} \bibinfo{author}{\bibfnamefont{S.~C.} \bibnamefont{Zhang}},
  \bibinfo{journal}{Science} \textbf{\bibinfo{volume}{314}},
  \bibinfo{pages}{1757} (\bibinfo{year}{2006}).

\bibitem[{\citenamefont{K{\"o}nig et~al.}(2007)\citenamefont{K{\"o}nig,
  Wiedmann, Br{\"u}ne, Roth, Buhmann, Molenkamp, Qi, and Zhang}}]{Koenig07}
\bibinfo{author}{\bibfnamefont{M.}~\bibnamefont{K{\"o}nig}},
  \bibinfo{author}{\bibfnamefont{S.}~\bibnamefont{Wiedmann}},
  \bibinfo{author}{\bibfnamefont{C.}~\bibnamefont{Br{\"u}ne}},
  \bibinfo{author}{\bibfnamefont{A.}~\bibnamefont{Roth}},
  \bibinfo{author}{\bibfnamefont{H.}~\bibnamefont{Buhmann}},
  \bibinfo{author}{\bibfnamefont{L.~W.} \bibnamefont{Molenkamp}},
  \bibinfo{author}{\bibfnamefont{X.~L.} \bibnamefont{Qi}}, \bibnamefont{and}
  \bibinfo{author}{\bibfnamefont{S.~C.} \bibnamefont{Zhang}},
  \bibinfo{journal}{Science} \textbf{\bibinfo{volume}{318}},
  \bibinfo{pages}{766} (\bibinfo{year}{2007}),
  \urlprefix\url{http://www.sciencemag.org/cgi/content/abstract/318/5851/766}.

\bibitem[{\citenamefont{Br\"une et~al.}(2012)\citenamefont{Br\"une, Roth,
  Buhmann, Hankiewicz, Molenkamp, Maciejko, Qi, and Zhang}}]{Bruene12}
\bibinfo{author}{\bibfnamefont{C.}~\bibnamefont{Br\"une}},
  \bibinfo{author}{\bibfnamefont{A.}~\bibnamefont{Roth}},
  \bibinfo{author}{\bibfnamefont{H.}~\bibnamefont{Buhmann}},
  \bibinfo{author}{\bibfnamefont{E.~M.} \bibnamefont{Hankiewicz}},
  \bibinfo{author}{\bibfnamefont{L.~W.} \bibnamefont{Molenkamp}},
  \bibinfo{author}{\bibfnamefont{J.}~\bibnamefont{Maciejko}},
  \bibinfo{author}{\bibfnamefont{X.-L.} \bibnamefont{Qi}}, \bibnamefont{and}
  \bibinfo{author}{\bibfnamefont{S.-C.} \bibnamefont{Zhang}},
  \bibinfo{journal}{Nat. Phys.} \textbf{\bibinfo{volume}{8}},
  \bibinfo{pages}{486} (\bibinfo{year}{2012}),
  \urlprefix\url{http://dx.doi.org/10.1038/nphys2322}.

\bibitem[{\citenamefont{Rothe et~al.}(2010)\citenamefont{Rothe, Reinthaler,
  Liu, Molenkamp, Zhang, and Hankiewicz}}]{Rothe10}
\bibinfo{author}{\bibfnamefont{D.~G.} \bibnamefont{Rothe}},
  \bibinfo{author}{\bibfnamefont{R.~W.} \bibnamefont{Reinthaler}},
  \bibinfo{author}{\bibfnamefont{C.-X.} \bibnamefont{Liu}},
  \bibinfo{author}{\bibfnamefont{L.~W.} \bibnamefont{Molenkamp}},
  \bibinfo{author}{\bibfnamefont{S.-C.} \bibnamefont{Zhang}}, \bibnamefont{and}
  \bibinfo{author}{\bibfnamefont{E.~M.} \bibnamefont{Hankiewicz}},
  \bibinfo{journal}{New J. of Physics} \textbf{\bibinfo{volume}{12}},
  \bibinfo{pages}{065012} (\bibinfo{year}{2010}).

\bibitem[{\citenamefont{Li and Tao}(2007)}]{Li07}
\bibinfo{author}{\bibfnamefont{Y.}~\bibnamefont{Li}} \bibnamefont{and}
  \bibinfo{author}{\bibfnamefont{R.}~\bibnamefont{Tao}},
  \bibinfo{journal}{Phys. Rev. B} \textbf{\bibinfo{volume}{75}},
  \bibinfo{pages}{075319} (\bibinfo{year}{2007}),
  \urlprefix\url{http://link.aps.org/doi/10.1103/PhysRevB.75.075319}.

\bibitem[{\citenamefont{Winkler and R\"ossler}(1993)}]{Winkler93}
\bibinfo{author}{\bibfnamefont{R.}~\bibnamefont{Winkler}} \bibnamefont{and}
  \bibinfo{author}{\bibfnamefont{U.}~\bibnamefont{R\"ossler}},
  \bibinfo{journal}{Phys. Rev. B} \textbf{\bibinfo{volume}{48}},
  \bibinfo{pages}{8918} (\bibinfo{year}{1993}),
  \urlprefix\url{http://link.aps.org/doi/10.1103/PhysRevB.48.8918}.

\bibitem[{\citenamefont{Schuurmans and 't~Hooft}(1985)}]{Schuurmans85}
\bibinfo{author}{\bibfnamefont{M.~F.~H.} \bibnamefont{Schuurmans}}
  \bibnamefont{and} \bibinfo{author}{\bibfnamefont{G.~W.}
  \bibnamefont{'t~Hooft}}, \bibinfo{journal}{Phys. Rev. B}
  \textbf{\bibinfo{volume}{31}}, \bibinfo{pages}{8041} (\bibinfo{year}{1985}),
  \urlprefix\url{http://link.aps.org/doi/10.1103/PhysRevB.31.8041}.

\bibitem[{\citenamefont{Shi et~al.}(2006)\citenamefont{Shi, Zhang, Xiao, and
  Niu}}]{ShiNiu06}
\bibinfo{author}{\bibfnamefont{J.}~\bibnamefont{Shi}},
  \bibinfo{author}{\bibfnamefont{P.}~\bibnamefont{Zhang}},
  \bibinfo{author}{\bibfnamefont{D.}~\bibnamefont{Xiao}}, \bibnamefont{and}
  \bibinfo{author}{\bibfnamefont{Q.}~\bibnamefont{Niu}},
  \bibinfo{journal}{Phys. Rev. Lett.} \textbf{\bibinfo{volume}{96}},
  \bibinfo{pages}{076604} (\bibinfo{year}{2006}),
  \urlprefix\url{http://link.aps.org/doi/10.1103/PhysRevLett.96.076604}.

\bibitem[{\citenamefont{K{\"o}nig et~al.}(2008)\citenamefont{K{\"o}nig,
  Buhmann, Molenkamp, Hughes, Liu, Qi, and Zhang}}]{Koenig08}
\bibinfo{author}{\bibfnamefont{M.}~\bibnamefont{K{\"o}nig}},
  \bibinfo{author}{\bibfnamefont{H.}~\bibnamefont{Buhmann}},
  \bibinfo{author}{\bibfnamefont{L.~W.} \bibnamefont{Molenkamp}},
  \bibinfo{author}{\bibfnamefont{T.}~\bibnamefont{Hughes}},
  \bibinfo{author}{\bibfnamefont{C.~X.} \bibnamefont{Liu}},
  \bibinfo{author}{\bibfnamefont{X.~L.} \bibnamefont{Qi}}, \bibnamefont{and}
  \bibinfo{author}{\bibfnamefont{S.~C.} \bibnamefont{Zhang}},
  \bibinfo{journal}{Journal of the Physical Society of Japan}
  \textbf{\bibinfo{volume}{77}}, \bibinfo{pages}{031007}
  (\bibinfo{year}{2008}),
  \urlprefix\url{http://jpsj.ipap.jp/link?JPSJ/77/031007/}.

\bibitem[{\citenamefont{M\"uhlbauer et~al.}(2013)\citenamefont{M\"uhlbauer,
  Budewitz, B\"uttner, Tkachov, Hankiewicz, Br\"une, Buhmann, and
  Molenkamp}}]{MuehlbauerbHankiewiczSupp}
\bibinfo{author}{\bibfnamefont{M.}~\bibnamefont{M\"uhlbauer}},
  \bibinfo{author}{\bibfnamefont{A.}~\bibnamefont{Budewitz}},
  \bibinfo{author}{\bibfnamefont{B.}~\bibnamefont{B\"uttner}},
  \bibinfo{author}{\bibfnamefont{G.}~\bibnamefont{Tkachov}},
  \bibinfo{author}{\bibfnamefont{E.~M.} \bibnamefont{Hankiewicz}},
  \bibinfo{author}{\bibfnamefont{C.}~\bibnamefont{Br\"une}},
  \bibinfo{author}{\bibfnamefont{H.}~\bibnamefont{Buhmann}}, \bibnamefont{and}
  \bibinfo{author}{\bibfnamefont{L.~W.} \bibnamefont{Molenkamp}},
  \bibinfo{journal}{arXiv:1306.2796}  (\bibinfo{year}{2013}).

\bibitem[{\citenamefont{Beenakker and van Houten}(1991)}]{BeenakkerHouten91}
\bibinfo{author}{\bibfnamefont{C.}~\bibnamefont{Beenakker}} \bibnamefont{and}
  \bibinfo{author}{\bibfnamefont{H.}~\bibnamefont{van Houten}},
  \textbf{\bibinfo{volume}{44}}, \bibinfo{pages}{1 } (\bibinfo{year}{1991}),
  ISSN \bibinfo{issn}{0081-1947},
  \urlprefix\url{http://www.sciencedirect.com/science/article/pii/S00811947086%
00910}.

\bibitem[{\citenamefont{Bardarson}(2008)}]{BardarsonPHD}
\bibinfo{author}{\bibfnamefont{J.~H.} \bibnamefont{Bardarson}},
  \emph{\bibinfo{title}{PhD Thesis}} (\bibinfo{publisher}{Leiden University},
  \bibinfo{year}{2008}).

\bibitem[{\citenamefont{Wimmer}(2008)}]{WimmerPHD}
\bibinfo{author}{\bibfnamefont{M.}~\bibnamefont{Wimmer}},
  \emph{\bibinfo{title}{PhD Thesis}} (\bibinfo{publisher}{Universit{\"a}t
  Regensburg}, \bibinfo{year}{2008}).

\bibitem[{\citenamefont{Rothe et~al.}(2012)\citenamefont{Rothe, Hankiewicz,
  Trauzettel, and Guigou}}]{Rothe12}
\bibinfo{author}{\bibfnamefont{D.~G.} \bibnamefont{Rothe}},
  \bibinfo{author}{\bibfnamefont{E.~M.} \bibnamefont{Hankiewicz}},
  \bibinfo{author}{\bibfnamefont{B.}~\bibnamefont{Trauzettel}},
  \bibnamefont{and} \bibinfo{author}{\bibfnamefont{M.}~\bibnamefont{Guigou}},
  \bibinfo{journal}{Phys. Rev. B} \textbf{\bibinfo{volume}{86}},
  \bibinfo{pages}{165434} (\bibinfo{year}{2012}),
  \urlprefix\url{http://link.aps.org/doi/10.1103/PhysRevB.86.165434}.

\bibitem[{\citenamefont{Datta}(2007)}]{Datta}
\bibinfo{author}{\bibfnamefont{S.}~\bibnamefont{Datta}},
  \emph{\bibinfo{title}{Electronic Transport in Mesoscopic Systems}}
  (\bibinfo{publisher}{Cambridge University Press}, \bibinfo{year}{2007}).

\bibitem[{\citenamefont{Liu et~al.}(2008)\citenamefont{Liu, Hughes, Qi, Wang,
  and Zhang}}]{LiuZhang08}
\bibinfo{author}{\bibfnamefont{C.}~\bibnamefont{Liu}},
  \bibinfo{author}{\bibfnamefont{T.~L.} \bibnamefont{Hughes}},
  \bibinfo{author}{\bibfnamefont{X.-L.} \bibnamefont{Qi}},
  \bibinfo{author}{\bibfnamefont{K.}~\bibnamefont{Wang}}, \bibnamefont{and}
  \bibinfo{author}{\bibfnamefont{S.-C.} \bibnamefont{Zhang}},
  \bibinfo{journal}{Phys. Rev. Lett.} \textbf{\bibinfo{volume}{100}},
  \bibinfo{pages}{236601} (\bibinfo{year}{2008}),
  \urlprefix\url{http://link.aps.org/doi/10.1103/PhysRevLett.100.236601}.

\bibitem[{\citenamefont{Liu et~al.}(2010)\citenamefont{Liu, Zhang, Yan, Qi,
  Frauenheim, Dai, Fang, and Zhang}}]{LiuZhang10}
\bibinfo{author}{\bibfnamefont{C.-X.} \bibnamefont{Liu}},
  \bibinfo{author}{\bibfnamefont{H.}~\bibnamefont{Zhang}},
  \bibinfo{author}{\bibfnamefont{B.}~\bibnamefont{Yan}},
  \bibinfo{author}{\bibfnamefont{X.-L.} \bibnamefont{Qi}},
  \bibinfo{author}{\bibfnamefont{T.}~\bibnamefont{Frauenheim}},
  \bibinfo{author}{\bibfnamefont{X.}~\bibnamefont{Dai}},
  \bibinfo{author}{\bibfnamefont{Z.}~\bibnamefont{Fang}}, \bibnamefont{and}
  \bibinfo{author}{\bibfnamefont{S.-C.} \bibnamefont{Zhang}},
  \bibinfo{journal}{Phys. Rev. B} \textbf{\bibinfo{volume}{81}},
  \bibinfo{pages}{041307} (\bibinfo{year}{2010}),
  \urlprefix\url{http://link.aps.org/doi/10.1103/PhysRevB.81.041307}.

\end{thebibliography}

\end{document}